\documentclass[longauth]{aa}

\usepackage{graphicx}
\usepackage{txfonts}
\usepackage[hidelinks,colorlinks=true,linkcolor=blue,citecolor=blue]{hyperref}
\usepackage[dvipsnames]{xcolor}
\usepackage{multicol}
\usepackage{multirow}
\usepackage{makecell}
\usepackage{cancel}

\usepackage{float}
\usepackage{latexsym}
\usepackage{amssymb}
\usepackage{amsmath}
\usepackage{amsfonts}
\usepackage{color}
\usepackage{cool}
\usepackage{subfig}
\usepackage{dcolumn}
\usepackage{soul}
\usepackage[utf8]{inputenc}
\usepackage[english]{babel}
\usepackage{import}
\usepackage{tabularx}
\usepackage{verbatim}
\usepackage{natbib}
\usepackage{tabu}
\usepackage{bm}

\bibpunct{(}{)}{;}{a}{}{,}

\newcommand{\hi}{H{\sc i}}

\makeatletter
\renewcommand*\aa@pageof{, page \thepage{} of \pageref*{LastPage}}
\makeatother
\begin{document} 
   
    \title {Testing synchrotron models and frequency resolution in BINGO 21 cm simulated maps using {\tt GNILC}}
    %\title {{\mathieu{21-cm?} Component Separation Tests for BINGO using {\tt GNILC}} \eduardo{jjzhang: shall we call it BINGO IX?}} 
    \author{Eduardo J. de Mericia\fnmsep
         \inst{1}
         \and
         {Larissa C. O. Santos}\thanks{larissa@yzu.edu.cn}
         \inst{2}
         \and
         {Carlos Alexandre Wuensche}
         \inst{1}
         \and
         {Vincenzo Liccardo}
         \inst{1}
         \and
         {Camila P. Novaes}
         \inst{1}
         \and
        {Jacques Delabrouille}
        \inst{3}
        \and
        {Mathieu Remazeilles}
        \inst{4}
        \and
        {Filipe Abdalla}
        \inst{5,6,7}
        \and
        {Chang Feng}
        \inst{8,9,10}
        \and
     %   {Elcio Abdalla}
     %   \inst{5}
     %   \and
        {Luciano Barosi}
        \inst{11}
        \and
        {Amilcar Queiroz}
        \inst{11}
        \and
        {Thyrso Villela}
        \inst{1,12,13}
        \and
        {Bin Wang}
        \inst{2,14}
        \and
        {Jiajun Zhang}
        \inst{15}
        \and
        {Andre~A.~Costa}
        \inst{2}
        \and
        {Elisa~G.M.~Ferreira}
        \inst{5,16}
        \and
        {Ricardo~G.~Landim}
        \inst{17}
          \and
         {Alessandro Marins}
        \inst{5}
        \and
        {Marcelo~V.~dos~Santos}
        \inst{11}
          }

\institute{
        %1
        Divis\~ao de Astrof\'isica, Instituto Nacional de Pesquisas Espaciais - INPE, Av. dos Astronautas 1758, 12227-010 - S\~ao Jos\'e dos Campos, SP, Brazil\\
        \email{eduardo.mericia@inpe.br}
        \and
        %2
        Center for Gravitation and Cosmology, YangZhou University, Yangzhou 224009, China\\
        \email{larissa@yzu.edu.cn}
        \and
        %3
        CNRS-UCB International Research Laboratory, Centre Pierre Binétruy, IRL2007, CPB-IN2P3, Berkeley, CA 94720, USA
        \and
        %4
        Instituto de F\'{i}sica de Cantabria (CSIC-UC), Avenida de los Castros s/n, 39005 Santander, Spain
        %Mathieu's affiliation 
        \and
        %5
        Instituto de F\'{i}sica, Universidade de S\~ao Paulo, R. do Mat\~ao, 1371 - Butant\~a, 05508-09 - S\~ao Paulo, SP, Brazil
        \and
        %6
        University College London, Gower Street, London, WC1E 6BT, UK
        \and
        %7
        Department of Physics and Electronics, Rhodes University, PO Box 94, Grahamstown, 6140, South Africa
        \and
        %8
        Department of Astronomy, School of Physical Sciences, University of Science and Technology of China, Hefei, Anhui 230026, China
        \and
        %9
        CAS Key Laboratory for Research in Galaxies and Cosmology, University of Science and Technology of China, Hefei, Anhui 230026, China
        \and
        %10
        School of Astronomy and Space Science, University of Science and Technology of China, Hefei, Anhui 230026, China
        \and
        %11        
        Unidade Acad\^emica de F\'{i}sica, Universidade Federal de Campina Grande, R. Apr\'{i}gio Veloso,  Bodocong\'o, 58429-900 - Campina Grande, PB, Brazil
        \and
        %12
        Centro de Gest\~ao e Estudos Estrat\'egicos SCS Qd 9, Lote C, Torre C S/N Salas 401 a 405, 70308-200 - Bras\'ilia, DF, Brazil
        \and
        %13
        Instituto de F\'{i}sica, Universidade de Bras\'{i}lia, Campus Universit\'ario Darcy Ribeiro, 70910-900 - Bras\'{i}lia, DF, Brazil
        \and
        %14
        School of Aeronautics and Astronautics, Shanghai Jiao Tong University, Shanghai 200240, China
        \and
        %15
        Shanghai Astronomical Observatory, Chinese Academy of Sciences, Shanghai 200030, China
        \and
        %16
        Kavli Institute for the Physics and Mathematics of the Universe (KIPMU), University of Tokyo, 5-1-5 Kashiwa-no-Ha, Kashiwa Shi, Chiba 277-8568, Kashiwa-shi, Japan
        \and
        %17
        Technische Universit\"at M\"unchen, Physik-Department T70, James-Franck-Strasse 1, 85748, Garching, Germany
}

   %\date{}

% \abstract{}{}{}{}{} 
% 5 {} token are mandatory
 
  \abstract
  % {} leave it empty if necessaryy 
  {
  The 21 cm hydrogen line is arguably one of the most powerful probes to explore the Universe, from recombination to the present times. To recover it, it is essential to separate the cosmological signal from the much stronger foreground contributions at radio frequencies. The \textbf{B}aryon Acoustic Oscillations from \textbf{I}ntegrated \textbf{N}eutral \textbf{G}as \textbf{O}bservations -- \textbf{BINGO} radio telescope is designed to measure the 21 cm line and detect baryon acoustic oscillations (BAOs)  using the intensity mapping technique.
  }
  {
  %The aim of this work is to build up on previous work from the BINGO collaboration \citep{Liccardo:2020_mission-simulations,Fornazier:2020} in order to test the analizes in light of using new and different foreground synchrotron models, as well as using ancillary data with different hardware configurations. The main idea of these tests is to optimize the foreground removal and recovery of the 21 cm signal across the full BINGO frequency band, as well as to determine an optimal number of frequency (redshift) bands for the signal recovery.
  %This work analyses the performance of the Generalized Needlet Internal Linear Combination ({\tt GNILC}) method combined with a power spectrum debiasing procedure, applied to a simulated BINGO mission, comparing two different synchrotron emission models and different instrumental configurations. The main idea of these tests is to optimize the foreground removal and recovery of the 21 cm signal across the full BINGO frequency band, as well as to determine an optimal number of frequency (redshift) bands for the signal recovery.
  This work analyses the performance of the Generalized Needlet Internal Linear Combination ({\tt GNILC}) method, combined with a power spectrum debiasing procedure. The method was applied to a simulated BINGO mission, building upon previous work from the collaboration. It compares two different synchrotron emission models and different instrumental configurations, in addition to the combination with ancillary data to optimize both the foreground removal and recovery of the 21 cm signal across the full BINGO frequency band, as well as to determine an optimal number of frequency (redshift) bands for the signal recovery.
  }
  {
  We have produced foreground emissions maps using the Planck Sky Model ({\tt PSM}), the cosmological \hi\ emission maps are generated using the  Full-Sky Log-normal Astro-Fields simulation Kit ({\tt FLASK}) package and thermal noise maps are created according to the instrumental setup. We apply the {\tt GNILC} method to the simulated sky maps to separate the \hi\ plus thermal noise contribution and, through a debiasing procedure, recover an estimate of the noiseless 21 cm power spectrum.
  }
  {
    %We found a near optimal reconstruction of the \hi\ signal using a 80 bins configuration, which resulted in a power spectrum reconstruction average error over all frequencies of 3\%. Furthermore, our tests showed that {\tt GNILC} is robust against different synchrotron emission models. Finally, adding an extra channel with CBASS foregrounds information to the 30 frequency bins configuration, we reduced the estimation error of the 21 cm signal from 7.2\% to 5.7\%.
    We found a near optimal reconstruction of the \hi\ signal using a 80 bins configuration, which resulted in a power spectrum reconstruction average error over all frequencies of 3\%. Furthermore, our tests showed that {\tt GNILC} is robust against different synchrotron emission models. Finally, adding an extra channel with CBASS foregrounds information, we reduced the estimation error of the 21 cm signal.
  }
  {
  %This optimisation of our earlier work in determining an optimal number of channels for binning the data impacts greatly the decisions about the hardware configuration before commissioning. We were able to recover the \hi\ signal with good efficiency in the harmonic space, but have yet to investigate the effect of $1/f$ noise in the data, which will possibly impact the recovery of the \hi\ signal. This issue will be addressed in a forthcoming work.
  %Determining an optimal number of channels for binning the data will greatly help the decisions about the hardware configuration before commissioning. We were able to recover the \hi\ signal with good efficiency in the harmonic space, but have yet to investigate the effect of $1/f$ noise in the data, which will possibly impact the recovery of the \hi\ signal. This issue will be addressed in a forthcoming work.  
  The optimisation of our previous work, producing a configuration with an optimal number of channels for binning the data, impacts greatly the decisions regarding BINGO hardware configuration before commissioning. We were able to recover the \hi\ signal with good efficiency in the harmonic space, but have yet to investigate the effect of $1/f$ noise in the data, which will possibly impact the recovery of the \hi\ signal. This issue will be addressed in a forthcoming work.
  }

  \authorrunning{E. J. de Mericia et al. }
%    \titlerunning{{\tt GNILC}}
  \titlerunning{Testing synchrotron models and frequency resolution for BINGO with {\tt GNILC}}

  \date{Received XX, XX, 2022; accepted XX, XX, 2022}
    
%   \keywords{ }

   \maketitle

   %\tableofcontents
 
%%%%%%%%%%%%%%%%%%%%%%%%%%%%%%%%%%%%%% NEW SECTION %%%%%%%%%%%%%%%%%%%%%%%%%%%%%%%%%%%%%%
\section{Introduction}
\label{intro}

Recently, it has been possible to study the creation and evolution of the Universe through various different cosmological experiments, shifting cosmology from a mere intellectual speculation to the prestigious nickname “precision cosmology”. Despite the sucessful endeavour of creating a cosmological model that matches most of the current observational challenges and constrain the cosmological parameter values $\sim 1\%$ precision, some aspects of the so-called “concordance model” need a much deeper investigation. Among them, the nature of the Dark Sector (dark matter and dark energy), which comprise about $95\%$ of the present composition of the Universe is among the most interesting open problems in modern astrophysics and cosmology. 

The \textbf{B}aryon Acoustic Oscillations from \textbf{I}ntegrated \textbf{N}eutral \textbf{G}as \textbf{O}bservations -- \textbf{BINGO} --  telescope, a unique instrument designed to be one of the first radiotelescopes to measure Baryon Acoustic Oscillations (BAOs) in the radio band, may unveil more details about the late evolution of the universe \citep{Abdalla:2021_BINGO-project}.

BINGO will detect the integrated signal of the cosmological neutral hydrogen (\hi\ signal) hyperfine transition at 1420 MHz (21 cm) frequency in redshift interval $0.127<z<0.449$, corresponding to a redshifted frequency interval $980<\nu<1260$ MHz. This survey will be conducted using a novel technique known as  intensity mapping (IM) \citep{Peterson:2006}. It allows the flux measurement of the signal produced by all \hi\ atoms, over large areas of the sky. Combined with the radial dimension offered by the observational frequency  band, IM can produce very large surveys, covering a significant fraction of the cosmological volume. Cosmological \hi\ datacubes can be used to probe the Universe at lower redshifts, complementing the cosmological information obtained by CMB temperature and polarization, weak lensing and SN Ia data. 

Nevertheless, the very faint radio signal from the 21 cm transition is easily covered by the much stronger foreground emission, mainly the Galactic diffuse component, constituted by synchrotron and free-free emissions, as well as extragalactic contribution from unresolved radio sources and CMB. Mitigating the foreground contamination to the detected signal is essential to recover the cosmological 21 cm information from the incoming data. 

Various methods to separate the astrophysical foreground components from the targeted cosmological signal are available in the literature, and most of them were already tested in the battlefield of CMB data analysis. A good component separation method greatly contributes to the accurate reconstruction of the 21 cm signal. On the one side, it works towards leaving the foreground contamination in the data to a minimum and, on the other side, prevents the loss of portions of the 21 cm data. An efficient component separation method (or a combination of complementary methods) reduces the uncertainties arising during the separation process and their propagation into the 21 cm signal power spectrum, avoiding introducing some bias in the estimation of cosmological parameters. 

Some of these methods assume prior knowledge of some foreground properties. Wiener Filtering \citep{Bunn:1994,Tegmark:1996}, for instance, assumes prior information on the frequency dependence of the foreground emission and of the power spectra of the components of sky emission \citep{Delabrouille:2007}.   Other approaches, such as the Gibbs sampling approaches, assume a parametric model for the emission laws of the sky components \citep{Jewell:2004,Wandelt:2004,Eriksen:2004,Larson:2007,Eriksen:2008}, and fit for the parameters and the amplitude of the various components in each sky pixel. These methods are considered to be ``non blind'' since they rely on prior information about the foregrounds that may not be very well understood. 

Alternatively, the so-called ``blind" approaches separate components of various physical origins in multi-frequency observations, relying only on their statistical independence. These methods include the Independent Component Analysis (ICA) \citep{Baccigalupi:2004}, which maximizes some measures of the non-gaussianity of independent sources. Examples are FastICA,  first used for foreground removal in CMB datasets \citep{Maino:2002b, Maino:2003}  and later applied to \hi\ mapping \citep{Wolz:2015}; and the Spectral Matching Independent Component Analysis (SMICA) \citep{Delabrouille:2003, Patanchon:2005, Betoule:2009},  which uses decorrelation to identify independent components.

The Internal Linear Combination (ILC), and its variants \citep{Tegmark:1996, Tegmark:2003, Bennett:2003, Eriksen:2004, Saha:2006, Delabrouille:2009, Basak:2012, Basak:2013,Remazeilles:2011}, are component separation techniques that have been extensively applied to CMB data analysis, in particular on WMAP and Planck survey data, to obtain foreground-cleaned CMB maps. This technique can be adapted to extract maps of other components, such as the 21 cm signal, using an extension called the Generalized Needlet ILC \citep{Remazeilles:2011b}, 
which will be used in this work to separate the 21 cm signal from the foregrounds. 

This work tests the performance of the {\tt GNILC} method in recovering the \hi\ signal in the presence of the different synchrotron models and with different binning in the frequency channels. We present reconstructed \hi\ simulated maps and power spectra in the presence of a combination of various  foregrounds and white noise. 

The paper is organized as follows. In Section \ref{sec:BINGO} we give a brief overview of the instrument. In Section \ref{sec:simulations}, we present the cosmological signal, astrophysical foregrounds and instrumental noise models used in our analysis, with a complementary discussion of the masking process and the description of our simulation plan. Section \ref{sec:GNILC} contains a brief description of {\tt GNILC}, the foreground removal method used in this work. Section \ref{sec:debiasing_procedure} describes the debiasing method used to correct the reconstructed power spectra. Then, in the Sections \ref{sec:results} and \ref{sec:conclusions}, we present the results and the conclusions of this work, respectively.

\section{Instrument overview}
\label{sec:BINGO}
 
The BINGO telescope is under construction in Paraiba, Brazil, and will be located at coordinates  7$^{\circ}$\,2$^\prime$\,27.6$^{\prime \prime}$\,S; 38$^{\circ}$\,16$^\prime$\,4.8$^{\prime \prime}$\,W. The site selection process is described in \cite{Peel:2019}. BINGO is designed to observe in a frequency interval between $980 \le \nu \le 1260 \,$ MHz, corresponding to a redshift interval $ 0.127 < z < 0.449$. BINGO will operate as a transit instrument, covering an instantaneous $\sim 15^{\circ}$-wide declination strip centered at $\delta = -15^{\circ}$. Its field of view will cover a $15.4^{\circ}$ wide declination strip, measured from center to center, with an angular resolution $\mathrm{FWHM \approx 40}$ arcmin at the central frequency of the band (1120 MHz). This accounts for a daily sky coverage of $5320$ square degrees. In this work, we consider $\mathrm{FWHM}=40'$ for the entire BINGO frequency range.

% Optica
The BINGO optics follows an off-axis crossed-Dragone configuration \citep{Dragone:1978}, with a primary, 40 m diameter paraboloid and a secondary 34 m diameter hyperboloid. The secondary dish illuminates a focal surface with 28 corrugated horns, each one feeding a polarizer coupled to two magic tees. The optics and the horn design and fabrication are described, respectively, in \citep{2021/abdalla_BINGO-optical_design, Wuensche:2020}. BINGO will operate with a full correlation receiver per horn, connecting two radiometer chains to each magic tee. Radiometers are expected to operate at a nominal system temperature $T_{\mathrm{sys}} \sim 70 $ K. 

Referring to the discussion of Section 2 of \cite{Wuensche:2021_instrument-description}, after 1 year of observations at $60\%$ duty cycle, with 30 frequency (redshift) bins and {\tt HEALpix} resolution $N_{\mathrm{side}}=128$ (comparable to the nominal angular resolution of the telescope) BINGO should achieve an estimated sensitivity of 102 $\mu$K.

\section{Simulated data}
\label{sec:simulations}

%The simulated data of this work were produced with the same tools used in \cite{Fornazier:2020}. However, the difference between the data used in the two works is that here we produce datasets with different numbers of frequency channels and different synchrotron models. Furthermore, we consider the thermal dust component in the data and do not include the cosmic microwave background (CMB), since it is very well characterized and could be directly removed from the data using Planck data. The maps are generated with the {\tt HEALpix} package \citep{Gorski:2005} with $N_{\mathrm{side}}=256$. More details on the components included in the simulated data are found in Section \ref{sec:components} and on simulation sets in Section \ref{sec:sets_of_simulations}.

This work builds upon and optimizes aspects of BINGO previous analyses. We use a simulated data set similar to the one described in \cite{Fornazier:2020}, and extends that work using different synchrotron models and assessing the capability of clearly recovering the \hi\ signal as a function of the number of channels used in the reconstruction.

This choice allows us to assess the level of systematic differences introduced by our lack of knowledge on the foregrounds. It also introduces different samplings of the data and includes ancillary data from the CBASS experiment \citep{cbass:2018_design_and_capabilities}. We particularly find that this will greatly help our decision regarding the final binning of the data, itself impacting the hardware configuration required by BINGO.

All maps used in this work are generated with the {\tt HEALpix} package \citep{Gorski:2005} with $N_{\mathrm{side}}=256$. More details on the components included in the simulated data are found in Section \ref{sec:components} and on simulation sets in Section \ref{sec:sets_of_simulations}.

\subsection{Cosmological, astrophysical and noise components}
\label{sec:components}

The \hi\ maps are generated using the Full-Sky Log-normal Astro-Fields Simulation Kit ({\tt FLASK}) package \citep{Xavier16}, using the $C_\ell$s created by the Unified Cosmological Library for $C_\ell$s ({\tt UCLCL}) code \citep{mcleod2017joint,Loureiro:2018qva}, and with a log-normal distribution with a very small deviation from a pure Gaussian. For more details on generating \hi\ maps using FLASK, see \cite{Liccardo:2020_mission-simulations}.

The foreground simulations are generated using the Planck Sky Model (PSM) software \citep{Delabrouille:2013}. The Galactic foregrounds are synchrotron radiation, free-free emission, the anomalous microwave emission (AME) and thermal dust (TD). The extragalactic contaminants include unresolved or faint radio point sources (FRPS) and the thermal and kinetic Sunyaev Zel'dovich (SZ) effects. Bright radio point sources are not considered, since they can be masked out during the analysis. For more details on foregrounds models, see \cite{Abdalla:2021_BINGO-project}, and generating their respective maps using {\tt PSM}, see \cite{Fornazier:2020}.

\begin{figure}[h]
\centering
    \includegraphics[width=0.45\textwidth]{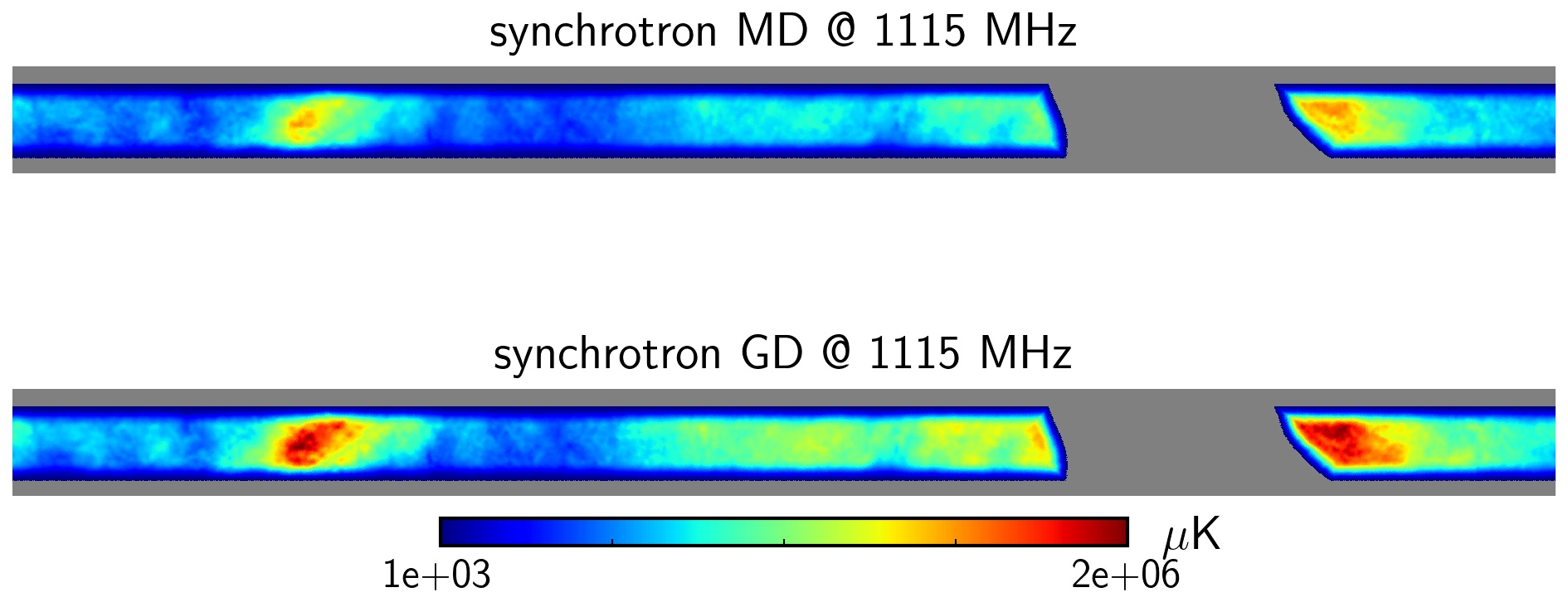}
    \caption{Synchrotron emission maps for the MD model ($Top$) and the GD model ($Bottom$), defined in the frequency channel centered at 1115 MHz and with a bandwidth of $\delta \nu$ $\sim$ 9.33 MHz (30 channels configuration).  The maps are presented in celestial coordinates and were convolved with a $\mathrm{FWHM}=40$ arcmin beam. The observed region corresponds to the BINGO covered area with the apodized galactic mask defined in the Section \ref{sec:mask}.} 
\label{fig:synchrotron_models_maps}
\end{figure}

\begin{figure}[h]
\centering
    \includegraphics[width=0.45\textwidth]{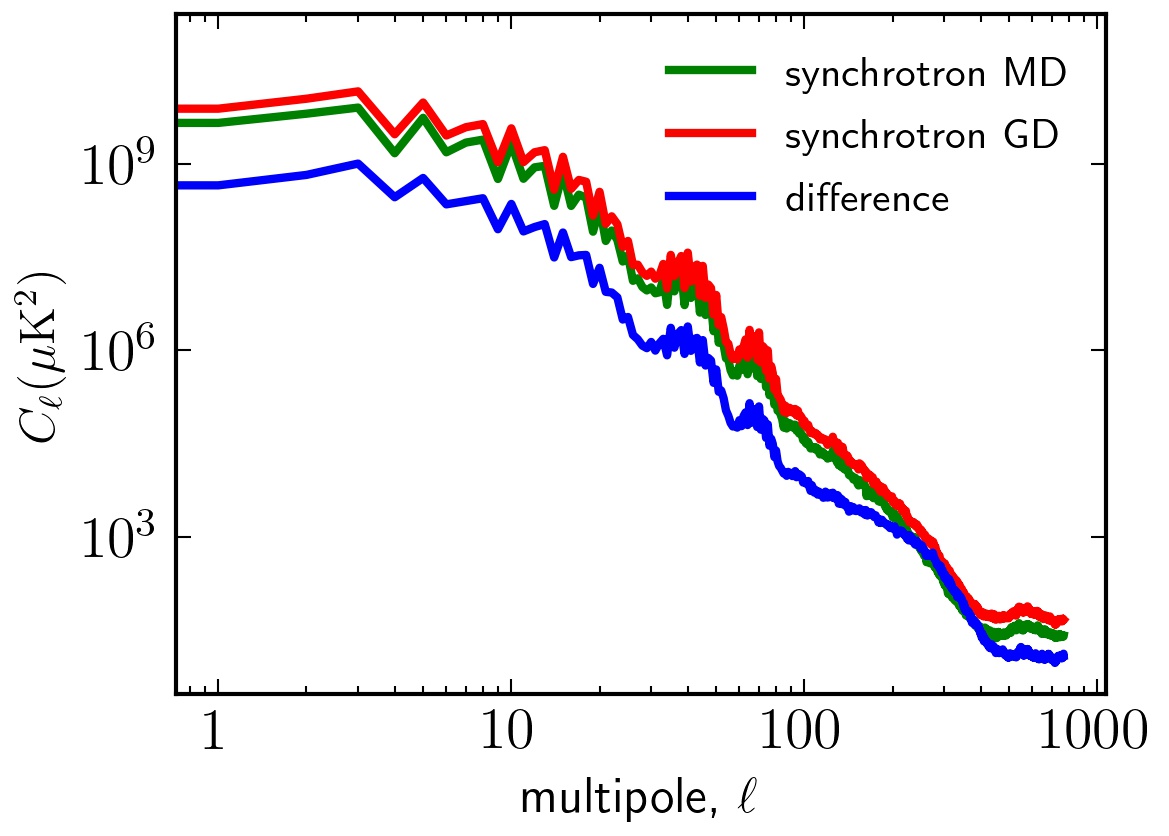}
    \caption{Power spectra referring to the MD (green) and GD (red) synchrotron maps shown in Figure \ref{fig:synchrotron_models_maps}, as well as the power spectrum calculated from the difference between them (blue), defined in the frequency channel centered at 1115 MHz and with a bandwidth of $\delta \nu$ $\sim$ 9.33 MHz (30 channels configuration). The maps were convolved with a $\mathrm{FWHM}=40$ arcmin beam. The observed region corresponds to the BINGO covered area with the apodized galactic mask defined in the Section \ref{sec:mask}.}
\label{fig:synchrotron_models_power_spectra}
\end{figure}

To produce the synchrotron component, we used as template the reprocessed Haslam 408 MHz all sky map, presented in \cite{Remazeilles:2015}. This map is then extrapolated to the BINGO frequencies (980-1260 MHz) through a power law, given by

\begin{equation}
\label{eq:synchrotron_power_law}
    T^{\rm{syn}}_{\nu}(p)  = T^{\rm{syn}}_{\nu_{0}}(p) \left[\frac{\nu}{\nu_{0}}\right]^{\beta_s(p)},
\end{equation}

\noindent where $T^{\rm{syn}}_{\nu_{0}}(p)$ is the template map, defined in the reference frequency $\nu_{0}$ and in the pixel $p$; $\beta_s(p)$ is the spatially variable spectral index. In our simulations, we considered synchrotron maps produced with two models of non-uniform $\beta_s$ over the sky: the \cite{Miville08} model (hereafter, synchrotron MD), which uses WMAP data at 23 GHz, and \cite{Giardino:2002} model (hereafter, synchrotron GD), which is the result of the combination of the full-sky map of synchrotron emission at 408 MHz from \cite{Haslam:1982}, the
northern-hemisphere map at 1420 MHz from \cite{Reich:1986} and the southern-hemisphere map at 2326 MHz from \cite{jonas1998rhodes}. The spectral index map of the synchrotron MD model has a mean value of -3.00 and a standard deviation of 0.06. These values for the synchrotron GD model are -2.9 and 0.1. Figure \ref{fig:synchrotron_models_maps} shows the synchrotron emission maps generated with the two pixel dependent spectral index models described above, and Figure \ref{fig:synchrotron_models_power_spectra} presents their respective power spectra. 

Although the thermal emission from dust grains is subdominant in the BINGO frequency band, we chose to include this component in the data in order to obtain a simulated sky closer to reality. We use the Galactic dust emission maps and a corresponding model obtained from the Planck 2015 data release using the {\tt GNILC} method \citep{Planck2016:dustemission}.

As instrumental systematics, we consider only the thermal (white) noise contribution over the sky, with an expected system temperature of 70 K. Due to the 28 horns arrangement in the BINGO focal plane (double rectangular scheme, see \cite{Wuensche:2021_instrument-description}), the observation time is not uniformly distributed over the sky area covered by the instrument. In this case, each horn covers the pixels of a fixed-latitude ring. Therefore, the total observation time of a pixel and, consequently, the root mean square (RMS) value of thermal noise per pixel depends on the latitude. Strong inhomogeneities in the innermost part of the covered region are avoided by repositioning the horns every year of the mission, within a total of $t_{\mathrm{obs}}$ = 5 years, which achieves a better distribution of observation time over the pixel in the covered area \citep{Liccardo:2020_mission-simulations}. Furthermore, we assume that the noise level map is the same for all frequency channels.

The RMS properties of the noise simulations used int his work follow the prescriptions published in \cite{Fornazier:2020}.

%Figure \ref{fig:noise_maps} shows the inhomogeneous RMS noise per-pixel ($\sigma_{\mathrm{pix}}$) map, produced with the characteristics described above; the white noise map, which is the result of multiplying the RMS white noise map by a Gaussian map with null mean and unitary standard deviation. It can be observed how the apodized mask (see Section \ref{sec:mask}) attenuates the noise in the region of the edges of the map, homogenizing the white noise map and favoring the reconstruction process of the 21 cm signal.

%\begin{figure*}[t]
%\centering
%\includegraphics[scale=0.75]{figures/noise_maps.jpg}
%\caption{White noise maps for the BINGO Phase 1 horn arrangement. The maps were generated with a $T_{\mathrm{sys}}$ = 70 K, 30 frequency bands, {\tt HEALpix} $N_{\mathrm{side}}=256$ and a fraction of the sky $f_ {\mathrm{sky}}=14.3\%$. $Left$: white noise level (RMS) map. $Center$: white noise map, generated using a Gaussian realisation of the corresponding RMS map (see text for details). $Right$: White noise realisation map with the apodized mask. The color scale is saturated at 3 times the RMS of a map with homogeneous coverage and same sky fraction. The maps are in celestial coordinates centered at $\alpha=0$ and $\delta=-18^{\circ}$.
%}
%\label{fig:noise_maps}
%\end{figure*}

\subsection{The masking process}
\label{sec:mask}

\begin{figure*}[h]
\centering
\includegraphics[scale=0.45]{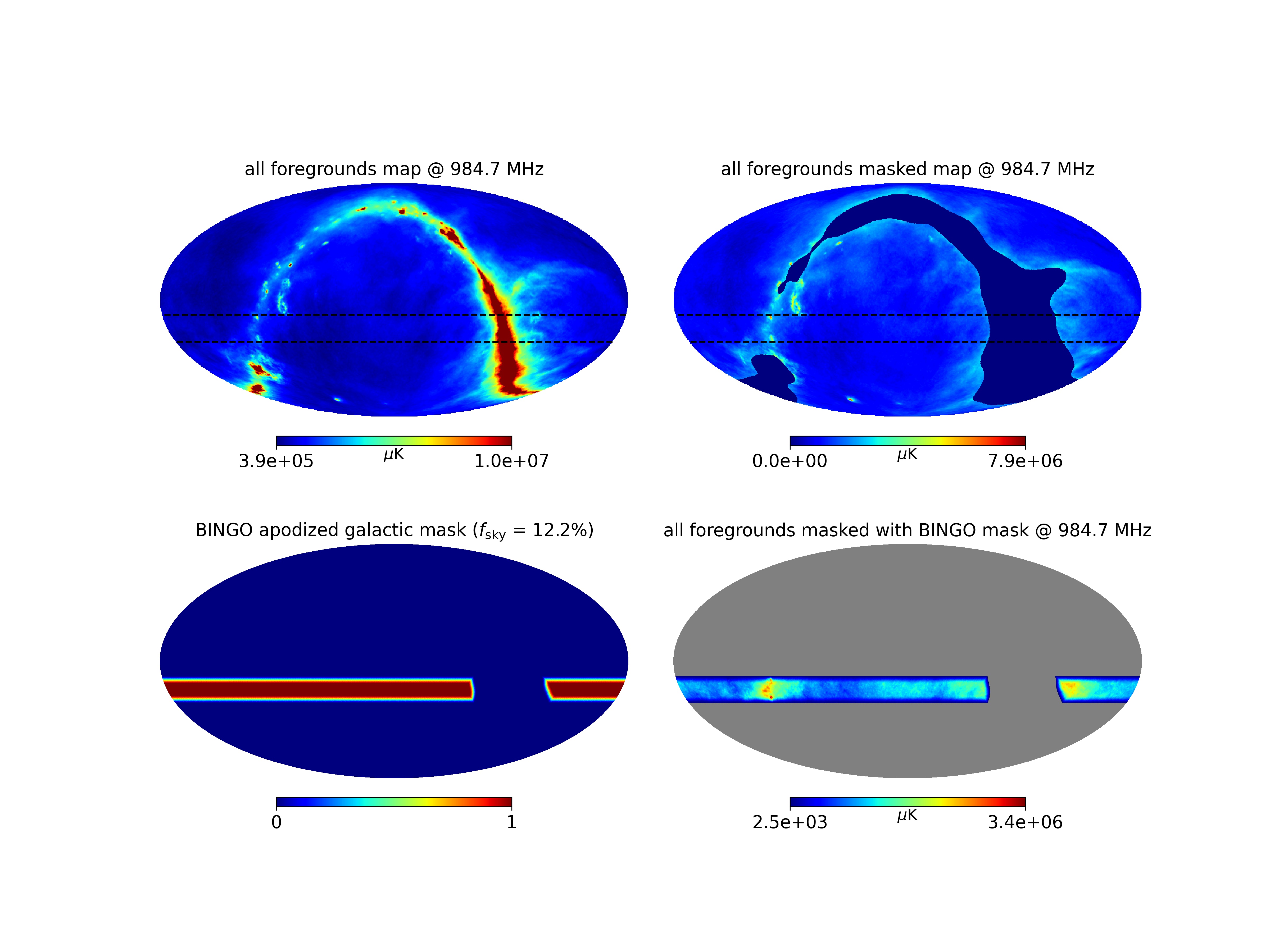}
\caption{Description of the masking process. Top left: map with the sum of all foregrounds considered in this work, including synchrotron MD model (see Section \ref{sec:components}), in the lowest frequency bin of a total of 30 channels, centered on 984.7 MHz and limited to 10 K. Top right: result of applying the galactic mask to the all foregrounds map. Bottom left: BINGO apodized (5 deg) galactic mask, preserving a sky fraction of 12.2\%. Bottom right: result of applying the BINGO mask to the all foregrounds map centered at 984.7 MHz. All maps are in celestial coordinates and the dashed lines delimit the BINGO covered area.}
\label{fig:galactic_mask}
\end{figure*}

In order to remove the region of the galactic plane where the foregrounds are more intense, and consequently facilitate the component separation process, we applied a galactic mask to the simulated maps. The Figure \ref{fig:galactic_mask} shows the combined emission at 984.7 MHz of all foregrounds considered in our simulations, as well as the region of the sky covered by BINGO. The observed region is a declination strip of $\approx 15^{\circ}$ centered on $\delta=-15^{\circ}$. Note that the region where the galactic foregrounds are more intense crosses the area covered by the instrument. The brightness temperature in this intersection region reaches 56 K.

We used a galactic mask which covers 20\% of the sky, cutting off the most intense emission region of the galactic plane. The result of applying this mask to the foregrounds map can be seen in the Figure \ref{fig:galactic_mask}. The maximum temperature within the area covered by BINGO after masking is about 6.5 K.

The BINGO observed region is defined by the feed horn arrangement in the focal plane and by the observation strategy \citep[see][]{Wuensche:2021_instrument-description}. The effective masked region is then the intersection between the galactic mask discussed above and the observed region. To avoid boundary artifacts in calculating the power spectrum of the maps, we use the {\tt NaMaster} \footnote{\url{https://namaster.readthedocs.io}} \citep{Alonso:2019} package to produce an apodization of type C2 and width $5^{\circ}$ in the mask. The result is shown in the bottom right panel of Figure \ref{fig:galactic_mask}, with a visible area of 12.2 \% of the sky.

% -------------------------------
\subsection{Assembling the simulated foreground maps}
\label{sec:sets_of_simulations}

The set of sky simulations used in this work was created to investigate the efficiency of our component separation pipeline against simulated foregrounds with different synchrotron models, for different numbers of frequency channels and and with the addition of foregrounds and noise data from another experiment, at a frequency outside the BINGO band.

In order to test the robustness of the method against different foregrounds models, we created two sets of simulated data. Each set contains a different synchrotron model (MD or GD), in addition to all other components described in the Section \ref{sec:components}. These sets were created with the BINGO project baseline configuration of 30 frequency bins and were called MD30 and GD30.

\begin{figure*}[h]
\centering
    \includegraphics[width=.40\textwidth]{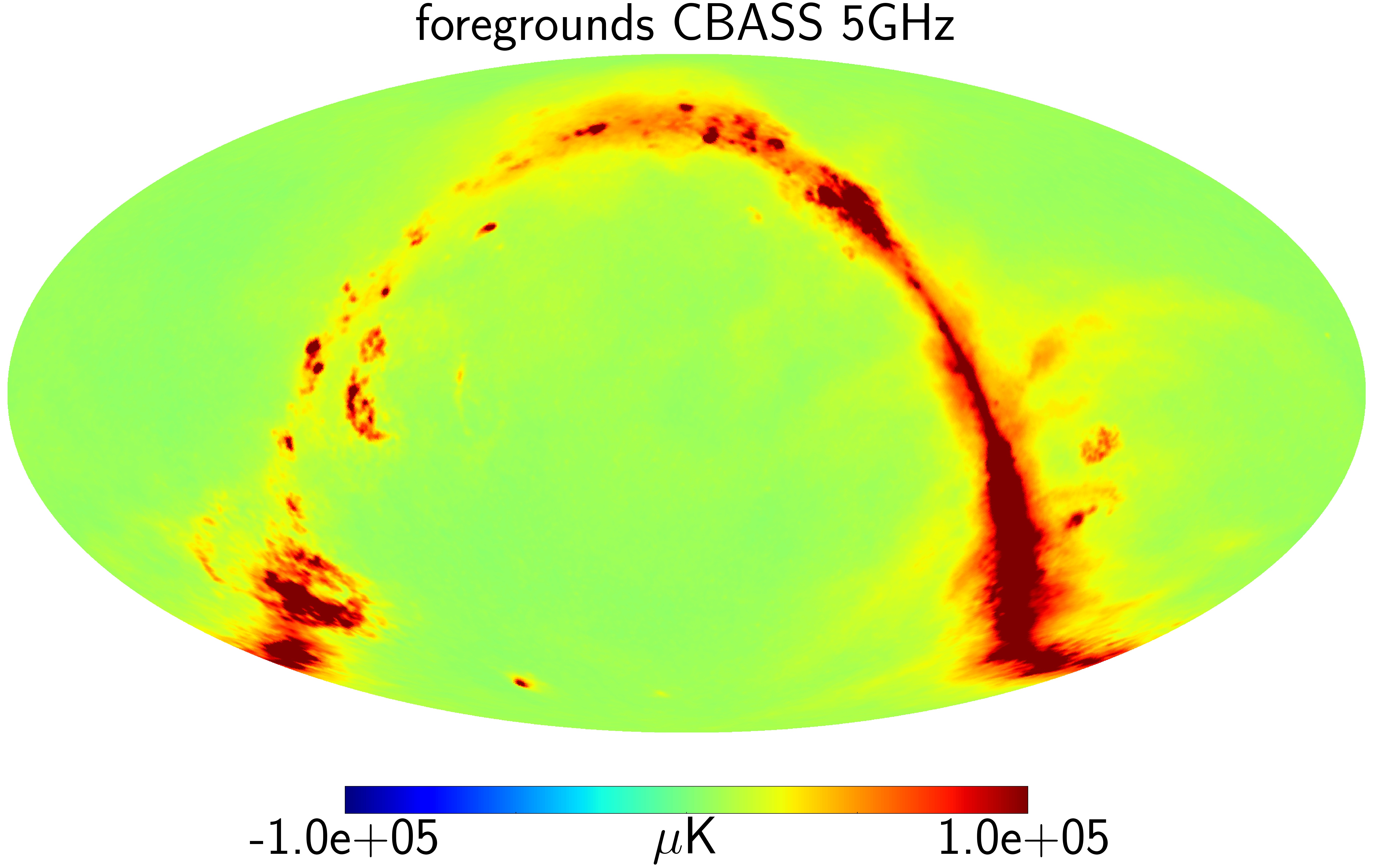}
    \includegraphics[width=.40\textwidth]{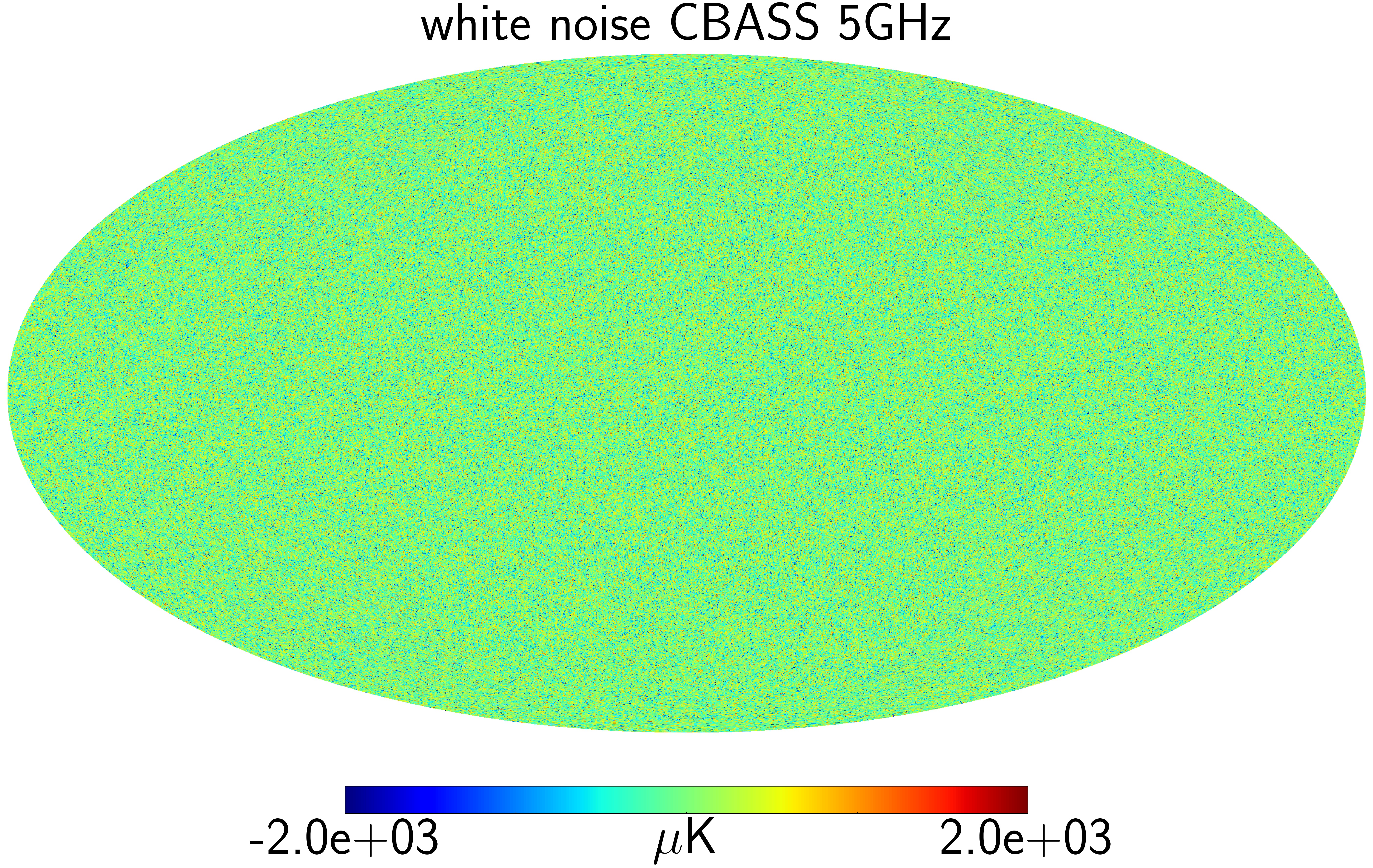}
    
    \caption{
    $Left$: CBASS all-sky foregrounds map, result of the sum of the components described in the Table \ref{tab:simulations_set}. $Right$: CBASS white noise map. The temperature scale of the foregrounds map is saturated at $\pm 10^5 \mu K$. The temperature scale of the noise map is saturated at $\pm 2 \times 10^3 \mu K$ and the noise level is $\sim 437 \mu K$.}
    \label{fig:cbass_maps}
\end{figure*}

The sky at low radio frequencies (< 10 GHz) is dominated by astrophysical foregrounds, mainly Galactic synchrotron emission. Independent foreground observations in frequencies outside the BINGO band may improve the characterization of the synchrotron contribution and other components, facilitating its removal. 
The C-Band All-Sky Survey (CBASS) is an all-sky survey at a frequency of 5 GHz and 1 GHz bandwidth, with a sensitivity $\lesssim 0.1 \mathrm{mK}$ RMS (per beam) and a resolution of 45 arcmin, designed to provide complementary data to the CMB surveys \citep{cbass:2018_design_and_capabilities}. Therefore, we also perform a component separation test adding the simulated CBASS sky and noise to the MD30 base dataset. We call the result of this data combination MD30+CBASS. The CBASS noise level is $\sim 437 \mu K$ for a {\tt HEALpix} $N_{\mathrm{side}}=256$. The simulated CBASS all-sky foreground and noise emission maps are shown in Figure \ref{fig:cbass_maps}.

% -------------------------------
\subsection{Simulation plan}
\label{sec:simulation_plan}

We generated sets of simulated data with 20, 30, 40, 60 and 80 frequency bins to evaluate the reconstruction efficiency the 21 cm signal with respect to different number of channels. To perform this test, we used the MD model for the synchrotron component, keeping the other components as described in the Section \ref{sec:components}. The created sets were called MD20, MD30, MD40, MD60 and MD80.

The BINGO input maps used in the tests described above are the result of the sum of \hi\ and foregrounds, convolved with a 40 arcmin beam, plus the estimated BINGO noise. In the case of combining BINGO and CBASS data, in addition to the BINGO frequency bins, we added a channel with CBASS foregrounds plus noise. In this configuration, the maps of the cosmological and astrophysical components, both from BINGO and CBASS, are convolved with a beam of 45 arcmin. The CBASS foregrounds map contains the same components considered in the BINGO data (see Section \ref{sec:components}). A summary of our simulation plan, defined by the foregrounds configuration, is tabulated in Table \ref{tab:simulations_set}.

\begin{table}[ht]
\caption{Foregrounds configurations and simulation plan}
\scriptsize
\centering
\begin{tabularx}{\columnwidth}{cccc}
%\begin{tabular}{|c|c|c|c|}
\hline
\\
{\bf \small Set} &  {\bf \small Foregrounds}  & \makecell{\bf \small Number of \\ \bf \small channels} \\[3ex]
%&   & {\bf \small Channels} \\[3ex]
\hline
\\
MD30      & Synchrotron MD + freefree + AME + TD + SZ + FPS & 30 (BINGO) \\[2ex]
\hline
\\
MD30      & Synchrotron MD + freefree + AME + TD + SZ + FPS & 30 (BINGO)\\        
        +CBASS &                                                 & + 1 (CBASS) \\[2ex]
\hline
\\
GD30      & Synchrotron GD + freefree + AME + TD + SZ + FPS & 30 (BINGO) \\[2ex]
\hline
\\
MD20      & Synchrotron MD + freefree + AME + TD + SZ + FPS & 20 (BINGO) \\[2ex]
\hline
\\
MD40      & Synchrotron MD + freefree + AME + TD + SZ + FPS & 40 (BINGO) \\[2ex]
\hline
\\
MD60     & Synchrotron MD + freefree + AME + TD + SZ + FPS & 60 (BINGO) \\[2ex]
\hline
\\
MD80      & Synchrotron MD + freefree + AME + TD + SZ + FPS & 80 (BINGO) \\[2ex]
\hline
\end{tabularx}
\label{tab:simulations_set}
\end{table}

% --------------------------------
\section{{\tt GNILC} Method}
\label{sec:GNILC}

The {\tt GNILC} method was developed to be used in CMB experiments \citep{Remazeilles:2011} and later adapted to the \hi\ IM  \citep{Olivari:2016}. The {\tt GNILC} basic idea is to use not only spectral information but also spatial information (angular power spectrum) to discriminate between foregrounds and our targeted signal, considered here as \hi\ plus thermal noise (see Section \ref{sec:21 cm+noise_reconstruction}). {\tt GNILC} is a blind method, that is, it does not assume prior knowledge about the properties of foregrounds, but only about the power spectrum of the desired signal. We present below a short summary of the method, as applied to this work, and refer the reader to the original papers cited above for more details.

The sky signal at the frequency $\nu$ and pixel $p$ can be modeled as

\begin{equation}
    \label{eq:sky_observation}
    d_{\nu}(p) = s_{\nu}(p) + f_{\nu}(p),
\end{equation}
where $s_{\nu}(p)$ is the true targeted signal and $f_{\nu}(p)$ is the contribution of the galactic and extragalactic foregrounds to the data. To adjust component separation to the local conditions of contamination, both in pixel and in harmonic spaces, the method uses a set of needlets  $h_{\ell}^{(j)}$, where $j$ indicates the range of angular scales selected by the bandpass. Thus, each $d_{\nu}(p)$ is decomposed into a map $d_{\nu}^{(j)}(p)$, which is the result of the inverse spherical harmonic transform of ${d_{\nu}(\ell,m) \times h_{\ell}^{(j)}}$. For each range of angular scales $j$, we compute the data covariance matrix at each pixel $p$, defining a pixel domain $N_p^{\left ( j \right )}$, centered on the pixel $p$, so that $N_{p}^{\left ( j \right )}$ is adjusted, for each needlet scale $j$, according to the choice of the ILC bias $b$ value. The bias $b$ and the set of needlets  $h_{\ell}^{(j)}$ are the input parameters for {\tt GNILC}.

{\tt GNILC} uses an estimate of the target signal covariance matrix, produced with a prior knowledge of the \hi\ $+$ noise power spectrum, to transform and diagonalize the data covariance matrix, disentangling foregrounds and signal subspaces on each needlet scale. This is the Principal Component Analysis (PCA)  \citep{Murtagh:1987_PCA} step of the method.

%{\tt GNILC} uses prior knowledge of the power spectrum of the targeted signal to produce signal map realisations, whose covariance matrix for each needlet scale contains separate subspaces for the sky emission and for the foregrounds. Following this step, a Principal Component Analysis (PCA)  \citep{Murtagh:1987_PCA} is performed on the diagonalized covariance matrix to disentangle foregrounds and target signal subspaces.

For each needlet scale $j$ and pixel $p$, {\tt GNILC} estimates the dimension $m$ of the foregrounds subspace (number of principal components) using the statistical Akaike Information Criterion (AIC) \citep{Akaike1974}. Thus, the method construct an ILC filter for each needlet scale, using the previous results, and apply it to the data, to obtain the reconstructed signal $\hat{s}_{\nu}^{(j)}(p)$.

%For each needlet scale $j$ and pixel $p$, we chose the appropriate eigenvectors of the data covariance matrix, using the statistical Akaike Information Criterion (AIC) \citep{Akaike1974} to select the principal components (foregorunds) of the covariance matrix. Thus, an ILC iteration is performed for each needlet scale to obtain the reconstructed signal $\hat{s}_{\nu}^{(j)}(p)$.

Finally, the reconstructed needlet maps, $\hat{s}_{\nu}^{(j)}(p)$ are transformed to spherical harmonic space and their harmonic coefficients are again band-pass filtered by the respective needlet window, $h_{\ell}^{(j)}$, and the filtered harmonic coefficients are transformed back to maps in pixel space,  creating back a single map per needlet scale. These maps are then added to give, for each frequency channel $\nu$, the full reconstructed map $\hat{s}_{\nu}(p)$.

{\tt GNILC} has two input parameters that control the locations used in the calculation of covariance matrices, both in the pixel and in the harmonic domains: the set of needlets and the ILC bias, and this work  tested {\tt GNILC} with different bias and needlet combinations. The best match for our simulations was the set of cosine-shaped needlets with peaks at $\ell=\left [ 0, 128, 384, 767 \right ]$ and the ILC bias $b=0.005$, and were the input parameters used in this work. The {\tt GNILC} method was tested in this work using simulated data with different configurations and the results are presented in Section \ref{sec:results}.

% -------------------------------

\section{Debiasing procedure}
\label{sec:debiasing_procedure}

Section \ref{sec:GNILC} describes the {\tt GNILC} method, referring to the signal of interest as the component to be recovered. Due to the characteristics of the 21 cm signal and the noise, we chose to recover both as a single component. For more details on this choice, see Section \ref{sec:21 cm+noise_reconstruction}. After obtaining the reconstructed 21 cm plus noise maps, we use a debiasing procedure to reconstruct the power spectrum of the 21 cm signal as a single component. This method consists of estimating the residual noise content and the loss of the \hi\ signal, after passing the data through the ILC filter (see Section \ref{sec:GNILC}), and correcting their effects on the power spectra of the {\tt GNILC} output maps. This procedure was also used by \cite{Fornazier:2020} in the reconstruction of 21 cm power spectra from simulated data in a configuration of 30 frequency bins. Here we explore the same method considering different configurations, as presented in Table \ref{tab:simulations_set}, and Section \ref{sec:21cm_power_spetrum_reconstruction}. A brief description of the technique is presented below.

The debiasing procedure to reconstruct the 21-cm power spectra is divided into two steps:
\begin{enumerate}
    \item Estimate the projected noise power spectra, $\hat{C}_{\ell}^{\mathrm{noise-proj},\nu}$, and debias the {\tt GNILC} map power spectra from this additive bias.
    \item Estimate the multiplicative bias, $\hat{b}_{\ell}^{\nu}$, and correct the noise-debiased {\tt GNILC} power spectra from it.
\end{enumerate}
These two steps can be summarized by the equation
\begin{equation}
    \label{eq:21cm_reconstructed_cl}
    \hat{C}_{\ell}^{\mathrm{21cm,\nu}} = \frac{C_{\ell}^{\mathrm{GNILC,\nu}}-\hat{C}_{\ell}^{\mathrm{noise-proj,\nu}}}{\hat{b}_{\ell}^{\nu}},
\end{equation}
where $\hat{C}_{\ell}^{\mathrm{21cm,\nu}}$ is our final estimate of the 21-cm power spectrum at the frequency channel $\nu$.

The additive noise bias $C_{\ell}^{\mathrm{noise-proj,\nu}}$ is estimated by generating $N_{\mathrm{realis}}$ white noise map realisations for each frequency channel $\nu$ and projecting them through the ILC weights matrix, computed in Section \ref{sec:GNILC} for the data. The power spectra of the resulting projected noise realisations are then averaged over all realisations.

The multiplicative bias $b_{\ell}^{\nu}$ is estimated by generating $N_{\mathrm{realis}}$ realisations of 21-cm signal maps at all frequency channels $\nu$ and computing the projected 21-cm signal by applying again the ILC weights matrix to the pure 21-cm map realisations. For each frequency channel $\nu$, we then compute the ratios between the power spectra of the projected 21-cm realisations and the power spectra of the input 21-cm realisations, which we average over all realisations in order to estimate $b_{\ell}^{\nu}$. 

The accuracy of the estimation of additive and multiplicative biases depends directly on the number of realizations used. However, the choice of $N_{\mathrm{realis}}$ is not free, but limited by the available computational capacity. The greater the number of channels or realisations used, the longer the debiasing processing time. Considering our computing resources and the settings adopted for the simulated data, we chose to test the debiasing procedure with two different numbers of realisations, as can be seen in Section \ref{sec:21cm_power_spetrum_reconstruction}.

% -------------------------------
\section{Results}
\label{sec:results}

Our component separation pipeline can be divided in two steps: the foreground removal stage, where we use {\tt GNILC} method, described in the Section \ref{sec:GNILC}, to recover the 21 cm plus noise signal from the BINGO simulated data maps; and the debiasing stage, where we use the procedure described in the Section \ref{sec:debiasing_procedure} to obtain the 21 cm power spectra from the {\tt GNILC} output maps. The BINGO simulated data maps used in this work are result of the sum of astrophysical foregrounds, \hi\ and thermal noise, as described in Section \ref{sec:simulations}. In the simulations that include an independent foreground observation, we add the CBASS 5 GHz map (foregrounds plus noise) to the set of BINGO simulated data maps. The results obtained in these two steps for the simulation plan in the Table \ref{tab:simulations_set} are described in the following subsections.

%Our component separation pipeline can be divided in two steps: the {\tt GNILC} stage, presented in the Section \ref{sec:GNILC}), where the 21 cm plus noise maps are recovered, and the debiasing stage, described in the Section \ref{sec:debiasing_procedure}), where the 21 cm power spectrum is reconstructed. In the first step, we start with the simulated data maps, a combination of foregrounds, \hi\ and thermal noise, as described in Section \ref{sec:simulations}, ending up with the recovered foreground clean \hi\ plus noise maps. When we add CBASS information, the extra channel is built by the sum of the respective foregrounds and noise maps. In this configuration, we reconstruct \hi\ plus noise in the BINGO channels and noise in the CBASS channel. In the debiasing step, we use $N_{\mathrm{realis}}$ realisations of \hi\ and white noise to estimate and remove the respective additive and multiplicative bias from the reconstructed 21 cm plus noise power spectra and recover the unbiased \hi\ power spectra. The results obtained in these two steps are divided into the two subsections below.

\subsection{21 cm plus noise maps reconstruction}
\label{sec:21 cm+noise_reconstruction}

When trying to recover the \hi\ signal as a single component using {\tt GNILC}, it was observed that the reconstructed power spectrum was contaminated by a residual content of thermal noise, mainly at small scales \citep{Olivari:2016}, which contributes to the 21 cm component estimation error. Since it is possible to obtain the thermal noise characteristics with good precision in an experiment, we chose to reconstruct the \hi\ and noise signals as a single component in a first step, as described in this section. Later, in Section \ref{sec:21cm_power_spetrum_reconstruction}, we use the debiasing procedure to estimate and remove the noise content present in our power spectra in order to be able to reconstruct the pure \hi\ signal. 

\begin{figure*}[h]
\centering
\includegraphics[scale=0.75]{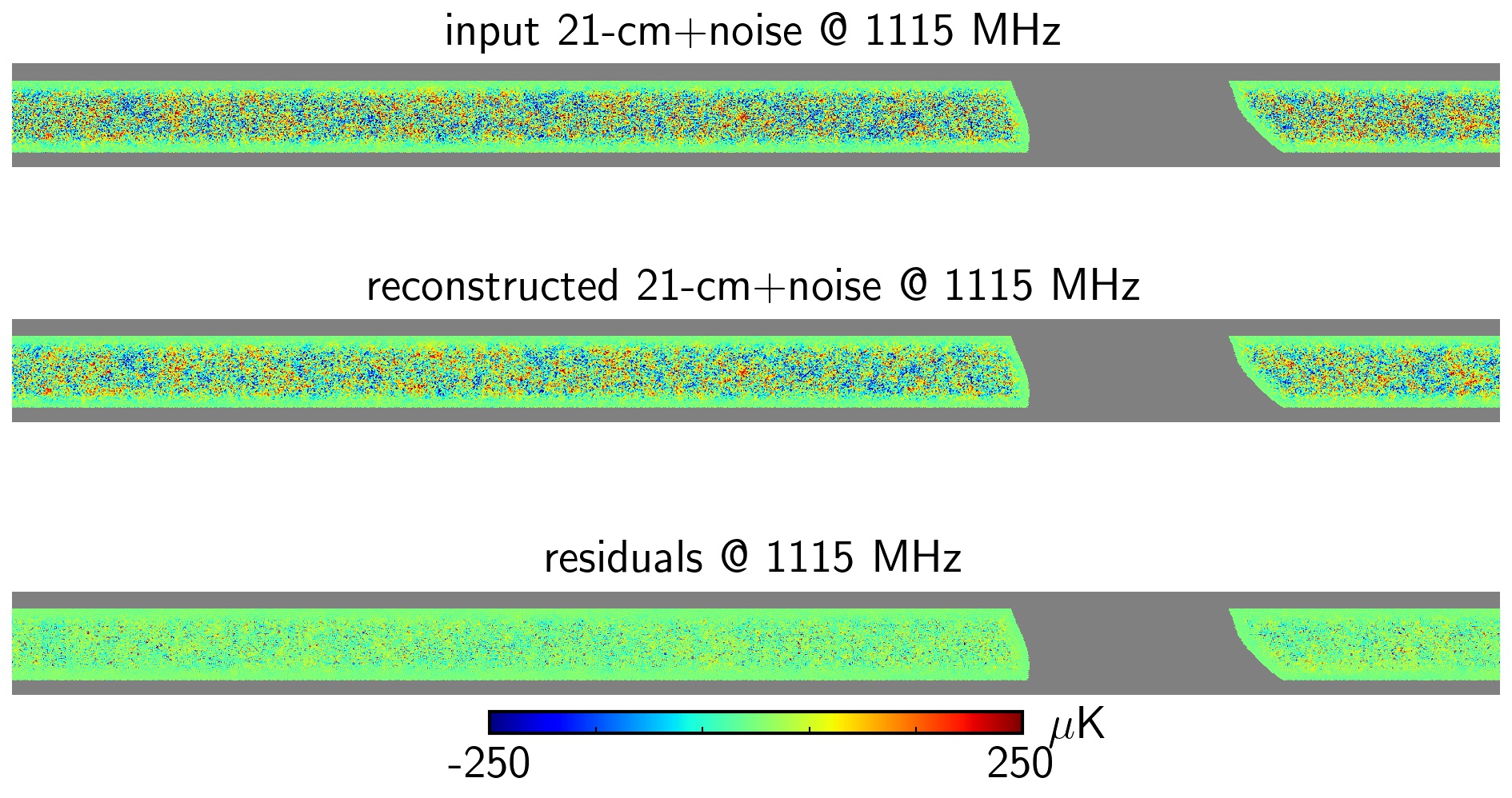}
\caption{Simulated 21 cm plus thermal noise input map ($Top$), {\tt GNILC} reconstructed map ($Middle$) and respective residuals, the difference between the two previous maps ($Bottom$), corresponding to a channel centered at 1115 MHz in MD30 configuration.
The maps are in celestial coordinates and are covered by the apodized galactic mask defined in Section \ref{sec:mask}. The \hi\ component was convolved with a $\mathrm{FWHM}=40$ arcmin beam.}  
\label{fig:hi_plus_noise_maps}
\end{figure*}

\begin{figure*}[h]
\centering
\includegraphics[scale=0.75]{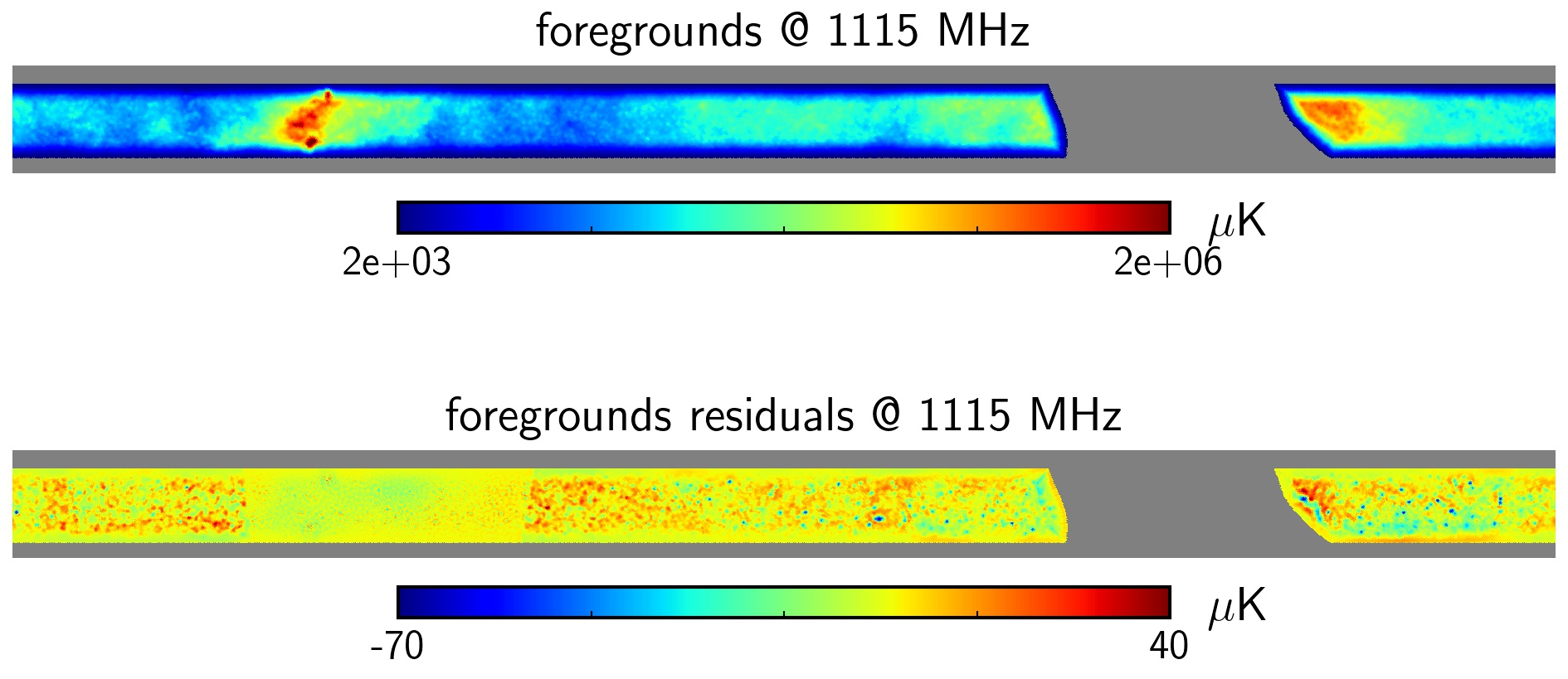}
\caption{Map containing the sum of all foregrounds described in Section \ref{sec:components} ($Top$) and their respective residuals after {\tt GNILC} ($Bottom$) at 1115 MHz in MD30 configuration (see Table \ref{tab:simulations_set}). The maps are in celestial coordinates and are covered with the apodized galactic mask defined in the Section \ref{sec:mask}. %Each map is a $15^\circ \times 15^\circ $ sky patch centered on $\alpha=0^{\circ}$ and $\delta=-18^{\circ}$, in celestial coordinates. 
Note the region of extreme declinations where the signal is attenuated by the apodized mask.
}
\label{fig:foregrounds_maps_full_strip}
\end{figure*}

\begin{figure*}[h]
\centering
\includegraphics[scale=0.65]{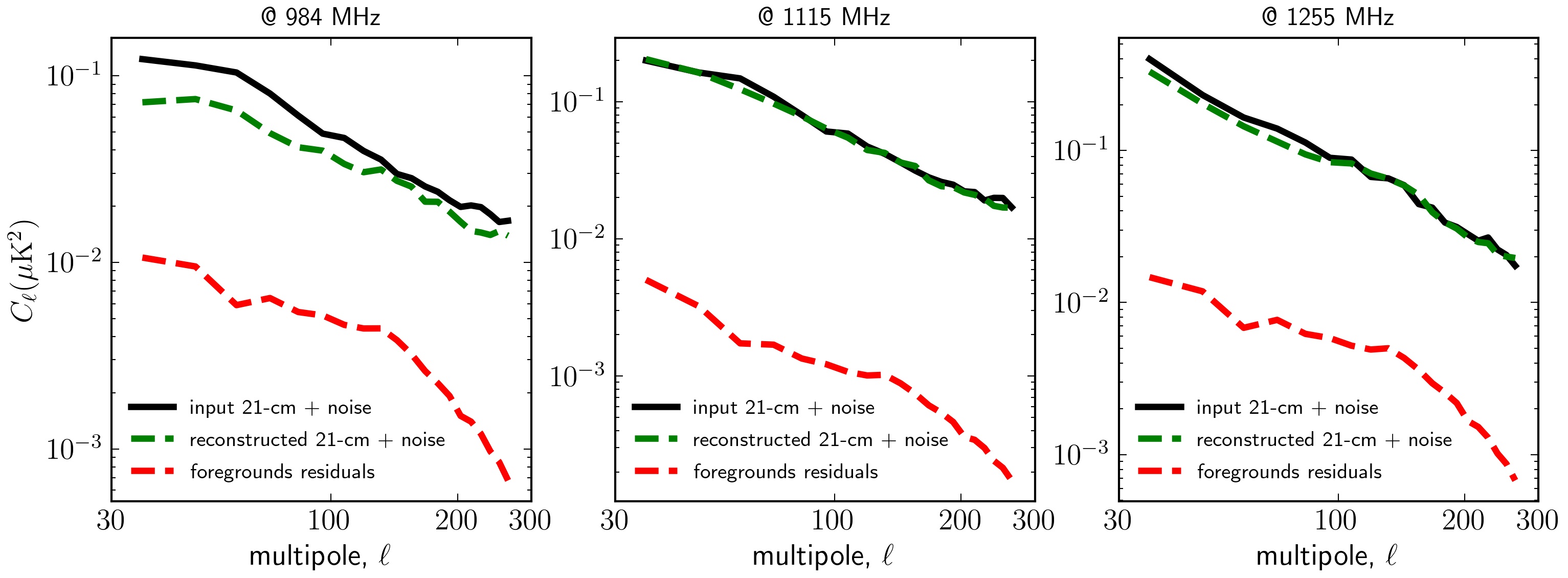}
\caption{Input 21 cm plus noise (black), reconstructed 21 cm plus noise (dashed green) and foregrounds residuals (dashed red) power spectra for frequency channels centered at 0.985 GHz ($left$), 1.115 GHz ($center$) and 1.255 GHz ($right$) for the MD30 configuration. The multipole range considered is $30 \leq \ell \leq 270$, with a multipole bin size of $\Delta \ell = 12$.}
\label{fig:gnilc_output_md30_power_spectra}
\end{figure*}

First, we present the results in the pixel domain for the MD30 case, the BINGO project baseline configuration. Figure \ref{fig:hi_plus_noise_maps} shows the input (expected), the reconstructed and the residual 21 cm plus noise maps, all observed near to the central frequency of the BINGO band (1120 MHz). Figure \ref{fig:foregrounds_maps_full_strip} shows the total foreground emission and the respective residual map after the component separation with {\tt GNILC}. Note the effect of the apodized galactic mask on the edges of the apparent region of the maps. This result shows the effectiveness of {\tt GNILC} in removing contaminants, since the content of foregrounds in the data is reduced from about $10^{5}$ to tens of $\mu$K. Figure \ref{fig:gnilc_output_md30_power_spectra} shows the comparison between the real and reconstructed power spectra of \hi\ plus noise, as well as the residual foregrounds after {\tt GNILC}. 

%%%%%%%%%%%%%%%%%%%%%%%%%%%

%%%%%%%%%%%%%%%%%%%%%%%%%%%%%%%%%%%%%%

The power spectra were plotted in the range of multipoles $30 \leq \ell \leq 270$, which is equivalent to angular scales $\sim 6^{\circ}$ to $\sim 0.7^{\circ}$. We do not consider very large angular scales, $\ell < 30$, due to the area of the sky covered by BINGO, and small angular scales, $\ell > 270^{\circ}$, due to the angular resolution of the instrument, 40 arcmin. Furthermore, as BINGO covers a fraction of the sky, the power spectra of its maps suffer from a loss of angular resolution, given by $\Delta \ell \sim 180^{\circ}/ \gamma_{\mathrm{max}}$, where $\gamma_{\mathrm{max}}$ is the maximum extent of the observed area \citep{ansari2010partial}. In our case, considering the repositioning of the horns during the mission, as described in \cite{Wuensche:2021_instrument-description} and \cite{Liccardo:2020_mission-simulations}, we have $\gamma_{\mathrm{max}}=17.5^{\circ}$. Thus, in order to respect this limitation and better adapt to the range of multipoles adopted ($30 \leq \ell \leq 270$), we choose to use a resolution of $\Delta \ell = 12$.

\begin{figure*}[h]
\centering
    \includegraphics[width=.45\textwidth]{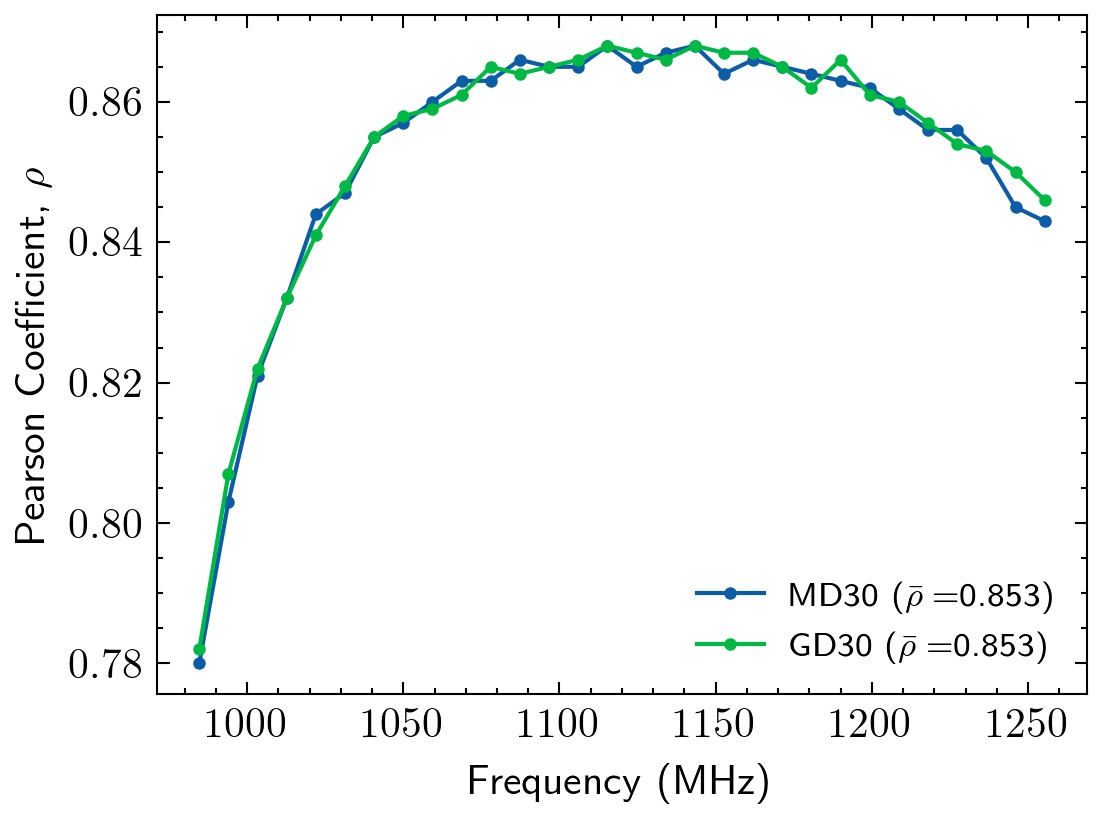}
    \includegraphics[width=.45\textwidth]{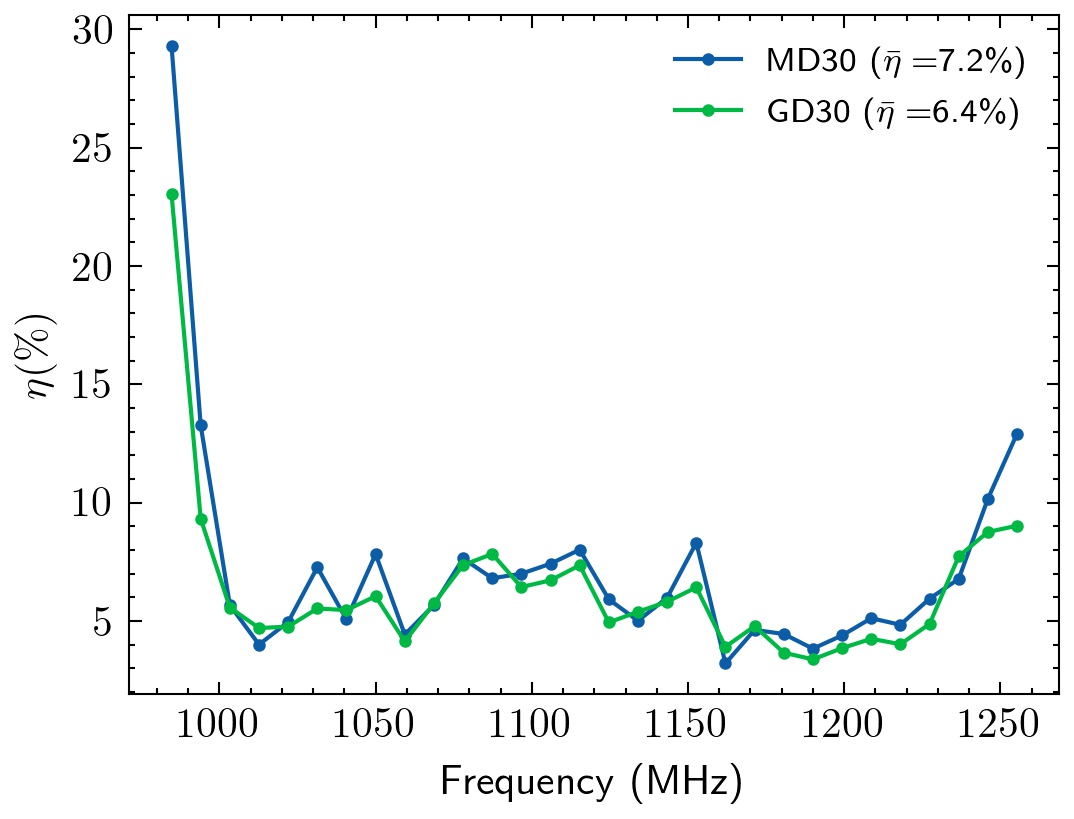}
    \\
    \includegraphics[width=.45\textwidth]{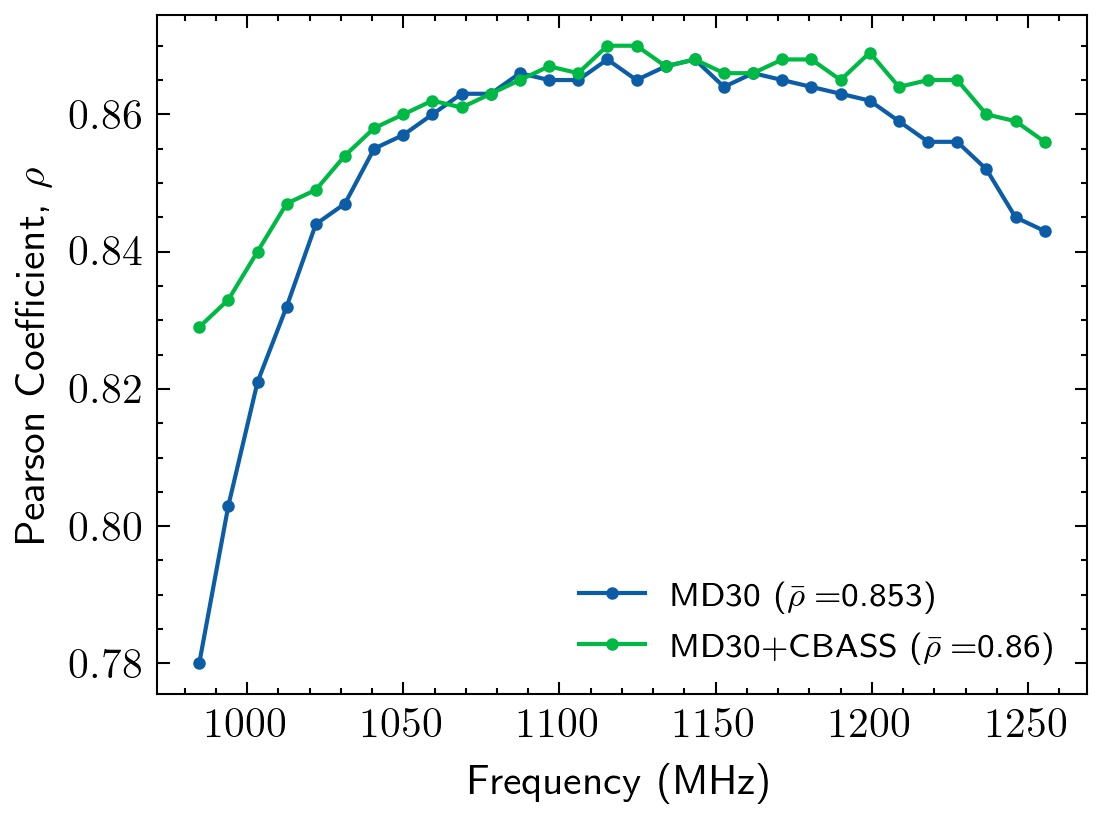}
    \includegraphics[width=.45\textwidth]{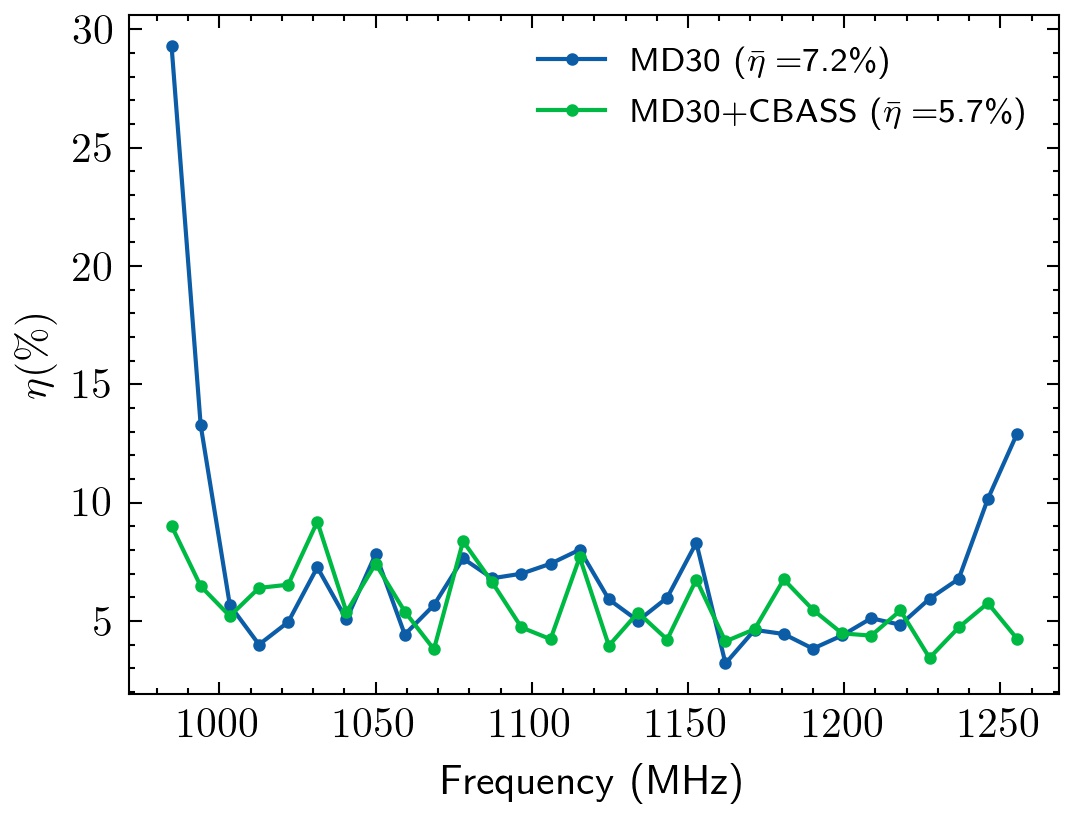}
    \\
    \includegraphics[width=.45\textwidth]{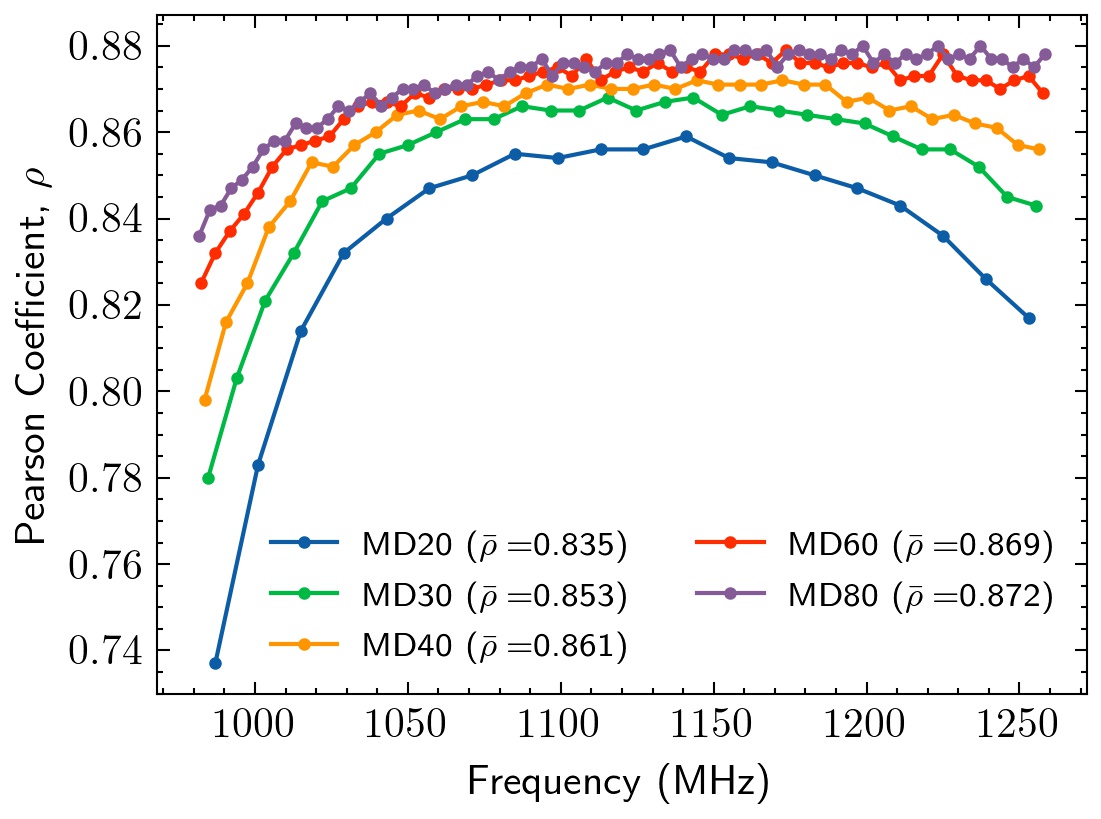}
    \includegraphics[width=.45\textwidth]{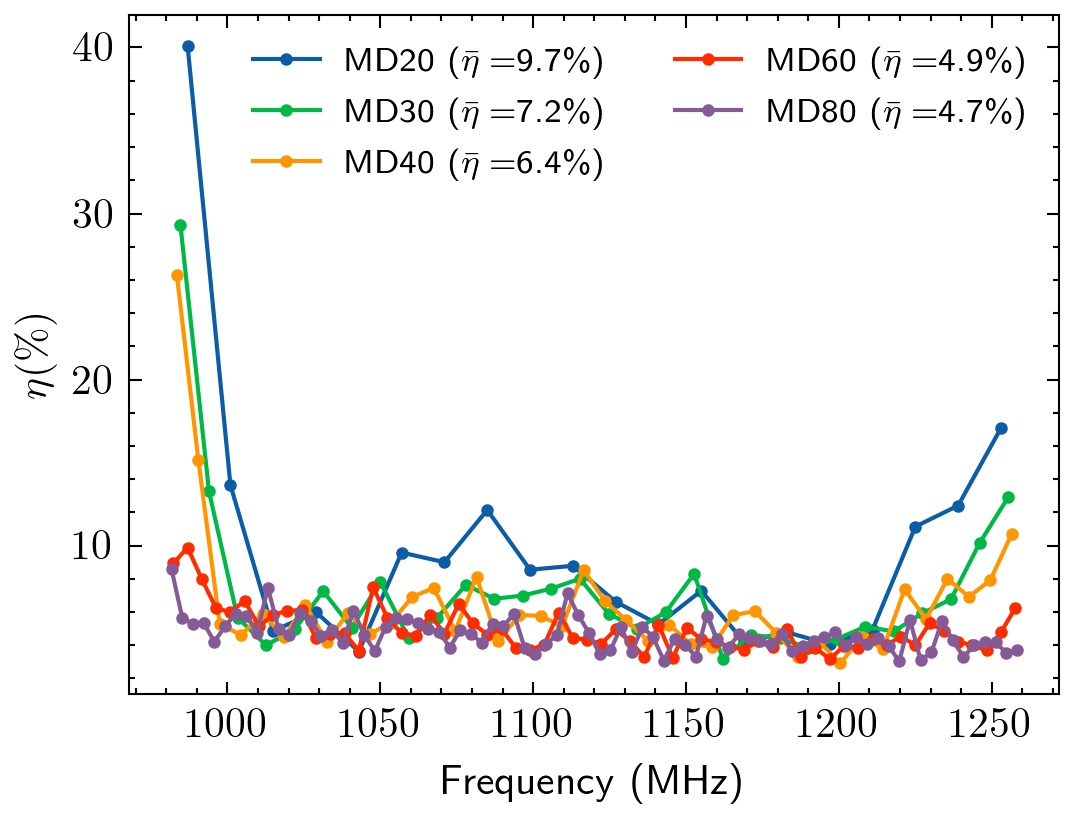}
    \caption{$Left$ $column$: Pearson coefficient $\rho$, calculated for each pair of expected and reconstructed 21 cm plus thermal noise maps. The $\bar{\rho}$ value in parentheses is the average error calculated over all frequency channels. $Right$ $column$: Reconstruction error of the 21 cm signal power spectrum $\eta$, calculated as an average over the multipole range $30 \leq \ell \leq 270$, on each frequency channel. The value $\bar{\eta}$ in parentheses is the average error calculated over all frequency channels. $Top$ $line$: Comparison between the results of $\rho$ and $\nu$ for the MD30 and GD30 configurations (different synchrotron models). $Center$ $line$: Comparison between the results of $\rho$ and $\nu$ for the MD30 and MD30+CBASS configurations (inclusion of an independent foreground observation). $Bottom$ $line$: Comparison between the results of $\rho$ and $\nu$ for the MD20, MD30, MD40, MD60 and MD80 configurations (different numbers of frequency bins).}
    \label{fig:compsep_pipeline_results}
\end{figure*}

We start checking the {\tt GNILC} response against different synchrotron models, the dominant component. For that, we performed the component separation with the MD30 and GD30 data sets (see Table \ref{tab:simulations_set}). To quantify the maps reconstruction efficiency, we used the Pearson coefficient, defined as

\begin{equation}
    \label{eq:pearson}
    \rho_{\nu} = \frac{\sum_{p}(\hat{s}_{\nu}(p)- \mu_{\hat{s}_{\nu}})(s_{\nu}(p)- \mu_{s_{\nu}})}{\sqrt{\sum_{p}(\hat{s}_{\nu}(p)- \mu_{\hat{s}_{\nu}})^{2}\sum_{p}(s_{\nu}(p)- \mu_{s_{\nu}})^{2}}} \ ,
\end{equation}

\noindent where $\hat{s}_ {\nu}(p)$ and $s_{\nu}(p)$ are the reconstructed and expected \hi\ plus noise maps, respectively, in a given pixel $p$ and frequency channel $\nu$. The $\mu_{\hat{s}_{\nu}}$ and $\mu_{s_{\nu}}$ terms are the mean value over the pixels of the reconstructed and expected \hi\ plus noise maps, respectively. The Pearson coefficients
between the input and reconstructed \hi\ plus noise maps are presented in Figure \ref{fig:compsep_pipeline_results}. It can be seen that the result varies very little between the two configurations. Furthermore, the average Pearson coefficient over all channels is the same for both cases, $\bar{\rho}=0.853$, which shows that {\tt GNILC} appears to be robust against different synchrotron models, in an analysis in the pixel domain. 

%\begin{figure*}[h]
%\centering
%    \includegraphics[width=.45\textwidth]{figures/pearson_md30_and_gd30.jpg}
%    \includegraphics[width=.45\textwidth]{figures/21cm_eta_md30_and_gd30.jpg}
%    \caption{$Left$: Pearson coefficients calculated for each pair of expected and reconstructed 21 cm plus thermal noise maps, in each frequency bin of the sets MD30 and GD30. $Right$: Reconstruction uncertainty $\eta$ of the 21 cm signal power spectrum, calculated as an average over the multipole range $30 \leq \ell \leq 270$, on each frequency channel of a dataset with different synchrotron models. The value $\bar{\eta}$ in parentheses is the average error over all channels in each dataset.}
%    \label{fig:md30_and_gd30_reconstruction}
%\end{figure*}

To verify the effect of adding an extra channel to the MD30 configuration, with the simulated data from CBASS experiment (see Section \ref{sec:sets_of_simulations}), we compared two cases: MD30 and MD30+CBASS. The Pearson coefficients between the {\tt GNILC} output maps and the expected \hi\ plus noise maps are plotted in Figure \ref{fig:compsep_pipeline_results}. As expected, we can notice an improvement in the reconstruction of \hi\ plus noise maps when we add the CBASS channel compared to the MD30 case. The average Pearson coefficient calculated across all channels is $\bar{\rho}=0.853$ without CBASS while is $\bar{\rho}=0.860$ with the addition of the CBASS channel. Furthermore, the Pearson coefficient shows a more significant improvement in the extreme frequencies of the band.

The accuracy with which {\tt GNILC} reconstructs the \hi\ plus noise map depends mainly on the number of frequency channels adopted. Theoretically, as we increase the number frequency bins, better {\tt GNILC} can remove astrophysical foregrounds from the data. The reason is that {\tt GNILC} is able to adapt the number of degrees of freedom $m$ (see Section \ref{sec:GNILC}), dedicated to describing the foregrounds, to the number of channels. This is done both in the pixel domain and in the harmonic domain, optimizing the \hi\ plus noise signal reconstruction \citep[see][]{Olivari:2016}. %However, as we will see in this section, in practice there is an upper limit to the number of channels, above which there is negligible improvement in the reconstruction of the signal of interest. 
Therefore, we tested the efficiency of the {\tt GNILC} in reconstructing 21 cm plus noise maps considering 20, 30, 40, 60 and 80 channels. For that, we used the datasets in the MD20, MD30, MD40, MD60 and MD80 configurations (see Table \ref{tab:simulations_set}).

The Pearson coefficients for configurations with different numbers of channels are shown in Figure \ref{fig:compsep_pipeline_results}. It is possible to observe that, in general, the Pearson coefficients values increase with the increase in the number of channels. However, the points seem to tend towards a limit. The difference in reconstruction efficiency improvement seems to decrease with increasing number of bins. Between the MD60 and MD80 configurations, this difference is already very small, indicating that there is a near optimal choice of the number of channels, which in our case appears to be around $N_{\mathrm{ch}}=80$. The average Pearson coefficient, calculated over all channels, is also directly related to the number of frequency bins. The calculated values range from $\bar{\rho}=0.835$, in the MD20 configuration, to $\bar{\rho}= 0.872$, in the MD80 case. As expected, an increase in the number of bins gives the method more degrees of freedom to describe and remove the foregrounds, improving the recovery of the \hi\ plus noise signal. However, there seems to be a limit in the number of channels, above which there is negligible improvement in the reconstruction of the signal of interest.

%\begin{figure*}[h]
%\centering
%    \includegraphics[width=.45\textwidth]{figures/pearson_md_all_nchannels.jpg}
%    \includegraphics[width=.45\textwidth]{figures/21cm_eta_md_all_frequencies.jpg}
%    \caption{$Left$: Pearson coefficients calculated for each pair of expected and reconstructed \hi\ plus thermal noise maps, in each frequency bin of the sets MD20, MD30, MD40, MD60 and MD80. $Right$: Reconstruction error $\eta$ of the 21 cm signal power spectrum, calculated as an average over a multipole range, defined here as $30 \leq \ell \leq 270$, on each frequency channel of a dataset with different numbers of channels (MD20, MD30, MD40, MD60 and MD80 cases, see Table \ref{tab:simulations_set}). The value $\bar{\eta}$ in parentheses is the average error over all channels in each dataset. It is possible to notice the error reduction with the increase in the number of channels.}
%    \label{fig:reconstruction_number_of_channels}
%\end{figure*}

\begin{figure*}[h]
\centering
\includegraphics[width=.45\textwidth]{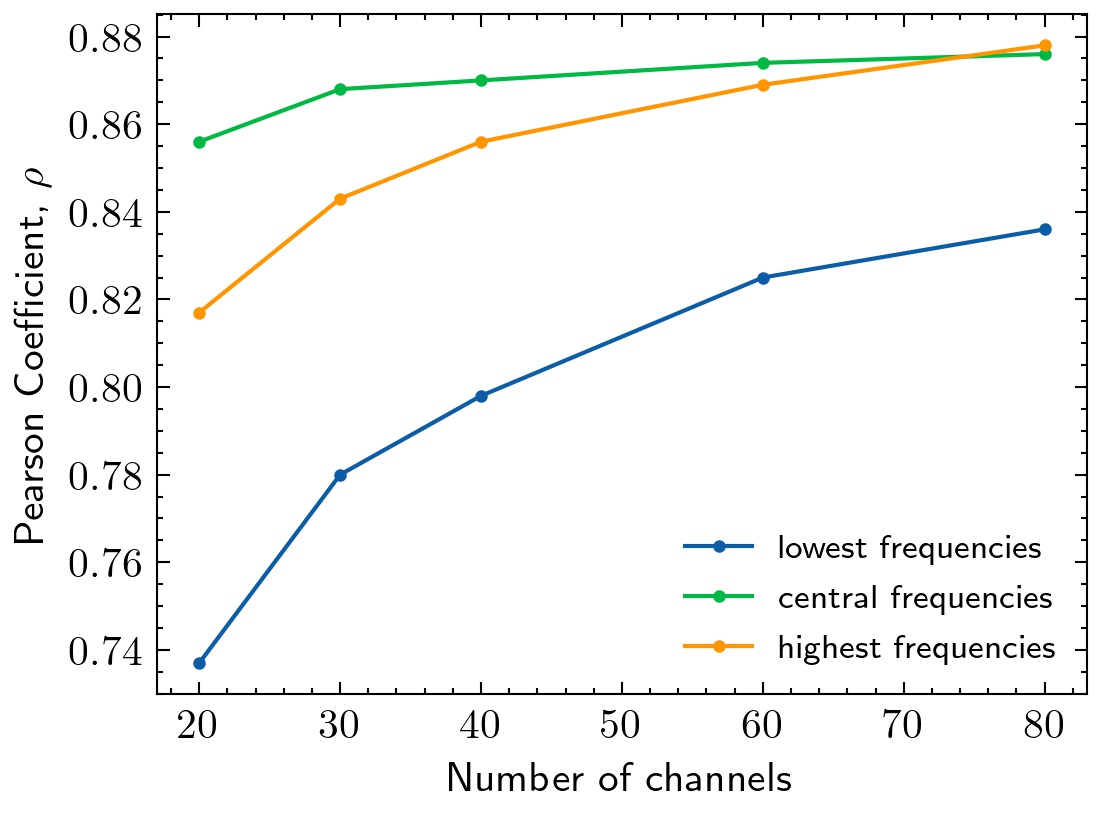}
\includegraphics[width=.45\textwidth]{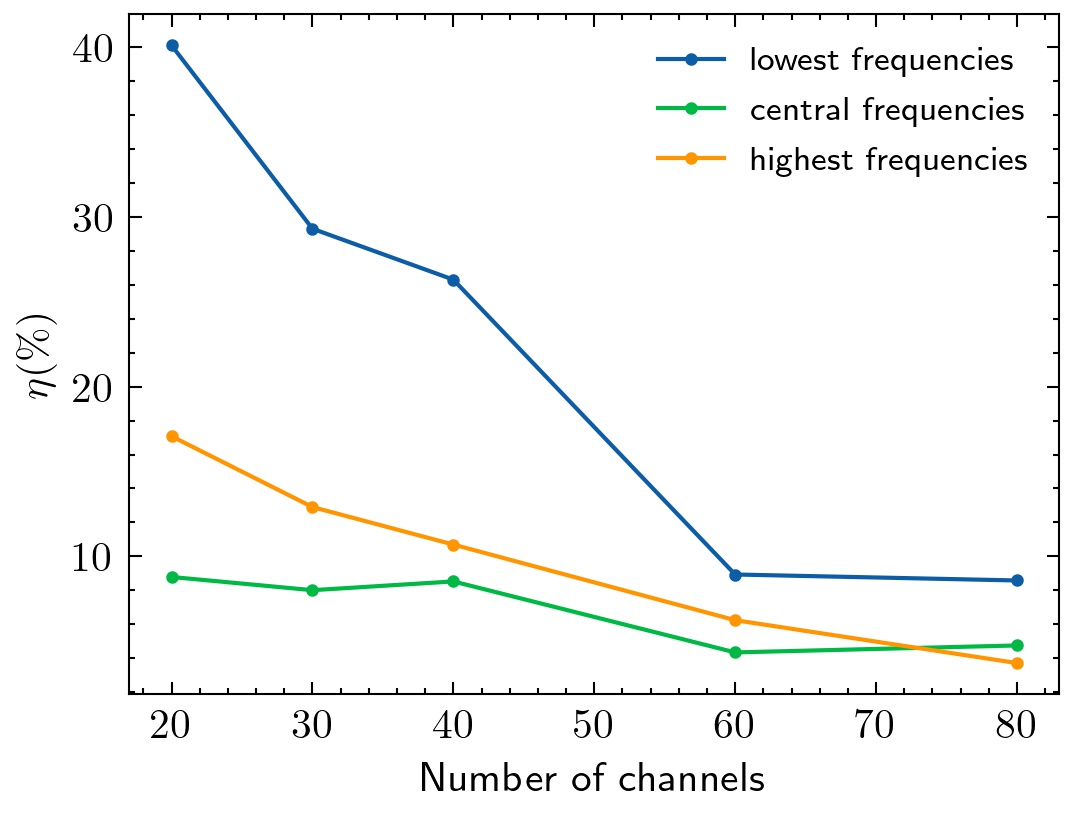}
\caption{$Left$: Pearson coefficients calculated for each pair of expected and reconstructed \hi\ plus thermal noise maps. $Right$: 21 cm power spectrum reconstruction error, calculated as an average over multipole range $30 \leq \ell \leq 270$. Both parameters are calculated for simulated data with different number of channels (MD20, MD30, MD40, MD60 and MD80 configurations), in channels centered at three different frequency channels: lowest frequencies (channels centered on 987, 984.7, 983.5, 982.3 and 981.7 MHz), central frequencies (channels centered on 1113, 1115.3, 1116.5, 1117.7 and 1118.2 MHz) and highest frequencies (channels centered on 1253, 1255.3, 1256.5, 1257.7 and 1258.2 MHz).}
\label{fig:pearson_per_number_of_channels}
\end{figure*}

Figure \ref{fig:pearson_per_number_of_channels} shows the Pearson coefficient as a function of the number of channels, calculated in channels centered at lower, central and higher frequencies, within the BINGO band (980-1260 MHz). In this figure it is possible to see better the stabilization trend in the efficiency of the reconstruction of the 21 cm plus noise maps with the increase in the number of channels. This effect is more evident in the central frequency channels, where the Pearson coefficients are already higher than in the other bins. This result reinforces what was said earlier regarding obtaining a near optimal reconstruction in the MD80 configuration.

So far we have presented the {\tt GNILC} performance in the reconstruction of the 21 cm plus noise maps. In the following section, we show the results of using the debiasing procedure, described in the Section \ref{sec:debiasing_procedure}, to recover the noiseless 21 cm power spectra.

\subsection{Noiseless 21 cm power spectrum reconstruction}
\label{sec:21cm_power_spetrum_reconstruction}

In this section, we present the results of noiseless 21 cm signal recovery in the multipole domain using the debiasing procedure presented in the Section \ref{sec:debiasing_procedure}. The power spectra of the {\tt GNILC} output maps ($C_{\ell}^{\mathrm{GNILC,\nu}}$), obtained in the previous section, are used here as one of the inputs for the bias correction method (see Equation \ref{eq:21cm_reconstructed_cl}). To measure the reconstruction error of the 21 cm power spectrum, we define the index $\eta(\nu)$, given by the absolute mean difference between the recovered $C_{\ell}^{\mathrm{21cm,r}}(\nu)$ and real $C_{\ell}^{\mathrm{21cm,s}}(\nu)$, normalized by the real power spectrum, according to

\begin{equation}
    \label{eq:eta_per_channel}
    \eta(\nu)=\frac{1}{N_{\ell}}\sum_{\ell=\ell_{\mathrm{min}}}^{\ell_{\mathrm{max}}}\left | \frac{C_{\ell}^{\mathrm{21cm,r}}(\nu)-C_{\ell}^{\mathrm{21cm,s}}(\nu)}{C_{\ell}^{\mathrm{21cm,s}}(\nu)} \right | \times 100 \% \ .
\end{equation}

\noindent This index is calculated for each frequency channel $\nu$ and for a multipole range %$\ell_{\mathrm{min}}\leq \ell \leq \ell_{\mathrm{max}}$
$30\leq \ell \leq 270$, as defined in Section \ref{sec:21 cm+noise_reconstruction}.

To perform the debiasing for all cases proposed in the Table \ref{tab:simulations_set} in a reasonable time, we chose to work with a low number of realizations. Therefore, we initially adopted a base number $N_{\mathrm{realis}}=10$ and after correcting the bias for all the configurations, we performed a test increasing this amount for the configuration with the best results.

Initially, we performed the debiasing on the power spectra of the {\tt GNILC} reconstructed maps, for the cases with different synchrotron models. Figure \ref{fig:compsep_pipeline_results} shows the error $\eta$ of power spectrum reconstruction of the 21 cm signal in each frequency channel of the MD30 and GD30 configurations. The mean reconstruction error for the case MD30 is 7.2\% and for the case GD30 is 6.4\%. This difference is due to the ability of {\tt GNILC} to adapt to the spatial variation of foregrounds to describe and remove them (see Section \ref{sec:GNILC}). Thus, maps generated with different models (MD and GD) may require different numbers of degrees of freedom to represent the contaminants in a given region of the sky, as well as to reconstruct the target signal. 

%\alex{Acho que a referencia a essa figura abaixo está errada, ela deveria ser a figura \ref{fig:foregrounds_power_spectra} atual.} 

\begin{figure*}[h]
\centering
\includegraphics[scale=0.65]{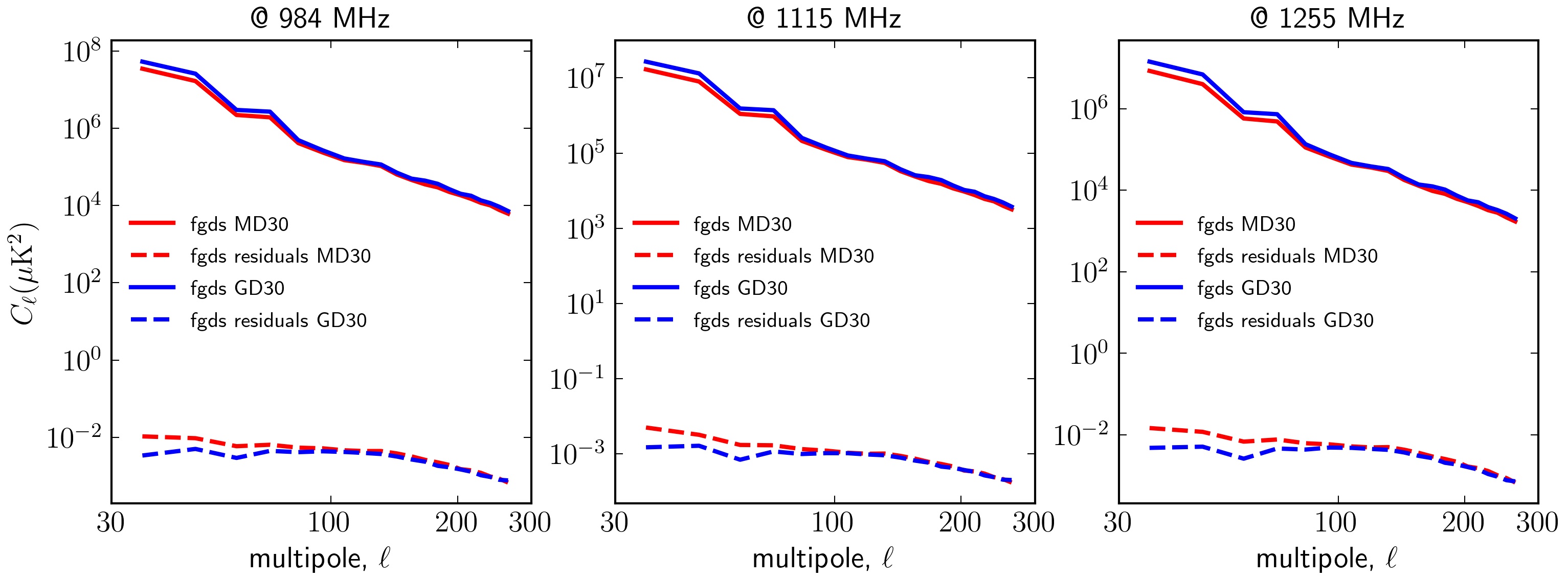}
\caption{Foregrounds power spectra in MD30 (red) and GD30 (blue) configurations, as well as their respective residuals present in the signal reconstructed with the {\tt GNILC} (dashed red and dashed blue). The power spectra are plotted for frequency channels centered at 0.985 GHz ($left$), 1.115 GHz ($center$) and 1.255 GHz ($right$). The multipole range considered is $30 \leq \ell \leq 270$, with a multipole bin size of $\Delta \ell = 12$.}
\label{fig:foregrounds_power_spectra}
\end{figure*}

Figure \ref{fig:foregrounds_power_spectra} shows the foregrounds power spectra in MD30 and GD30 configurations and their respective residuals after component separation with {\tt GNILC}. It is possible to notice that at scales $\ell<100$, {\tt GNILC} removes more foregrounds from the GD30 configuration than from the MD30. Furthermore, this residuals are preserved after the debiasing step, %according to the Equations \ref{eq:gnilc_output_power_spectra} and \ref{eq:21cm_reconstructed_cl}, 
contributing to the error in estimating the power spectrum of the 21 cm signal.

Next, we present the results of the debiasing procedure, considering the addition of a channel with simulated CBASS map to the BINGO data, in its basic project configuration (MD30 case). Figure \ref{fig:compsep_pipeline_results} shows the error $\eta$ of power spectrum reconstruction of the 21 cm signal in each frequency channel. The reconstruction error for the case MD30 is 7.2\% and for the case MD30+CBASS is 5.7\%. As expected, the inclusion of an extra channel with CBASS data increased the number of degrees of freedom available to describe the foregrounds, improving the reconstruction of the 21 cm \hi\ signal.

Figure \ref{fig:compsep_pipeline_results} shows the reconstruction error $\eta$ of the 21 cm signal in the harmonic domain, in each frequency channel, considering simulated data with different numbers of channels: MD20, MD30, MD40, MD60 and MD80. An improvement in the cosmological signal reconstruction can be observed with the increase in the number of channels because the number of dimensions available to describe the components increases, as discussed in Sections \ref{sec:GNILC} and \ref{sec:21 cm+noise_reconstruction}. The smallest average error over all channels is $\bar{\eta}=4.7\%$, obtained with 80 frequency channels. 

Figure \ref{fig:pearson_per_number_of_channels} shows the power spectrum reconstruction error, calculated in three different channels, corresponding to the lowest, central and highest frequencies, within the BINGO band (980-1260 MHz). In this figure, it is possible to notice a reduction in the estimation error with the increase in the number of channels, in addition to a stabilization trend of this reduction. This effect is more evident in the lowest frequency channels, where the estimation error are already higher for smaller numbers of channels. This result corroborates what was presented in Section \ref{sec:21 cm+noise_reconstruction}, regarding obtaining a near optimal reconstruction with 80 frequency bins.

\begin{figure}
\centering
\includegraphics[width=8.5cm]{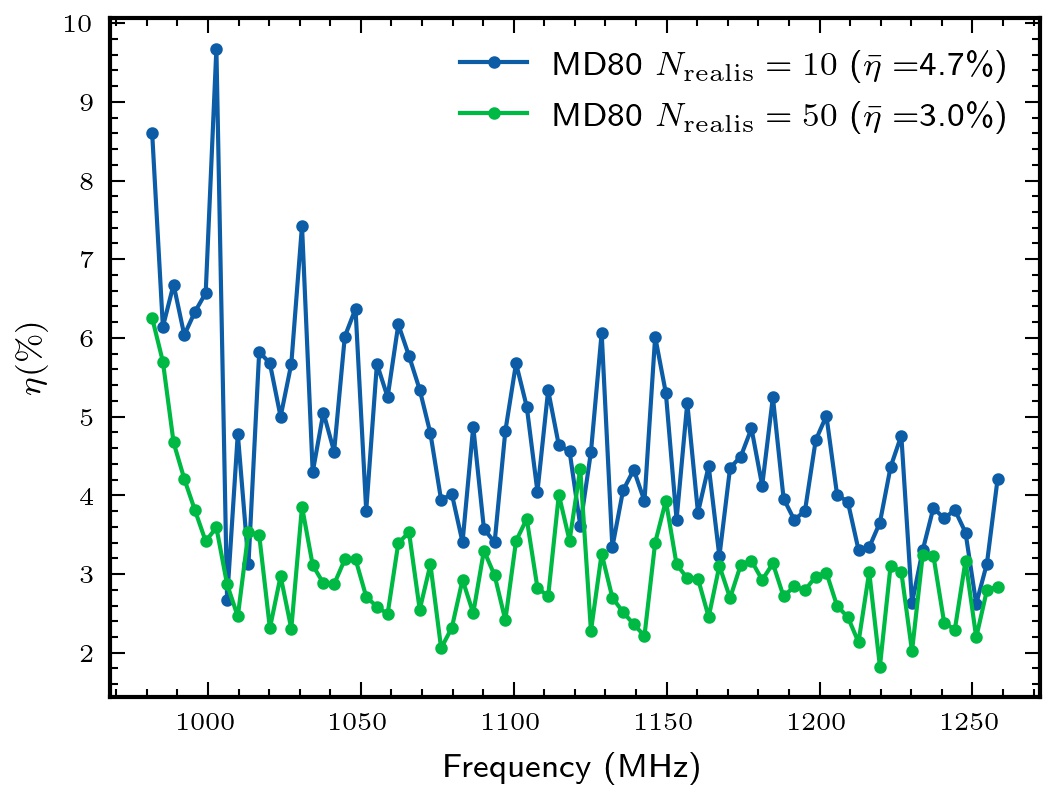}
\caption{Reconstruction error $\eta$ of the 21 cm signal power spectrum, calculated as an average over a multipole range, defined here as $30 \leq \ell \leq 270$, on each frequency channel of a dataset with MD80, considering two different number of realisations ($N_{\mathrm{realis}}=10$ and $N_{\mathrm{realis}}=50$). The value $\bar{\eta}$ in parentheses is the average error over all channels in each dataset. The mean error for the $N_{\mathrm{realis}}=10$ case is $\bar{\eta}=4.7\%$ and for the $N_{\mathrm{realis}}=50$ case is $\bar{\eta}=3.0\%$}
\label{fig:eta_different_nrealisations}
\end{figure}

To evaluate the effect of varying the number of realisations on the 21 cm spectrum reconstruction error, we reprocessed the debiasing procedure with a greater number of \hi\ and noise realisations. For that, we chose the MD80 configuration, the case with the best results in the previous analysis, and performed the bias correction with $N_{\mathrm{realis}}=50$ realisations. Then, we compared with the previous results, as presented in the Figure \ref{fig:eta_different_nrealisations}. The 21 cm signal reconstruction error for the case MD80 with $N_{\mathrm{realis}}=10$ is 4.7\% and with $N_{\mathrm{realis}}=50$ is 3.0\%. As expected, the reconstruction of the 21 cm signal is better when we use more \hi\ and thermal noise realisations to estimate the additive and multiplicative bias present in the {\tt GNILC} reconstructed 21 cm plus noise power spectrum. The reason is that with more realizations, the method improves the estimation of the reconstructed spectra bias. Our objective here is not to optimize the number of realizations, which would require much more computational processing time than we had available, but to perform a sensitivity analysis of the debiasing method with this parameter.

\begin{figure*}[h]
\centering
\includegraphics[scale=1.5]{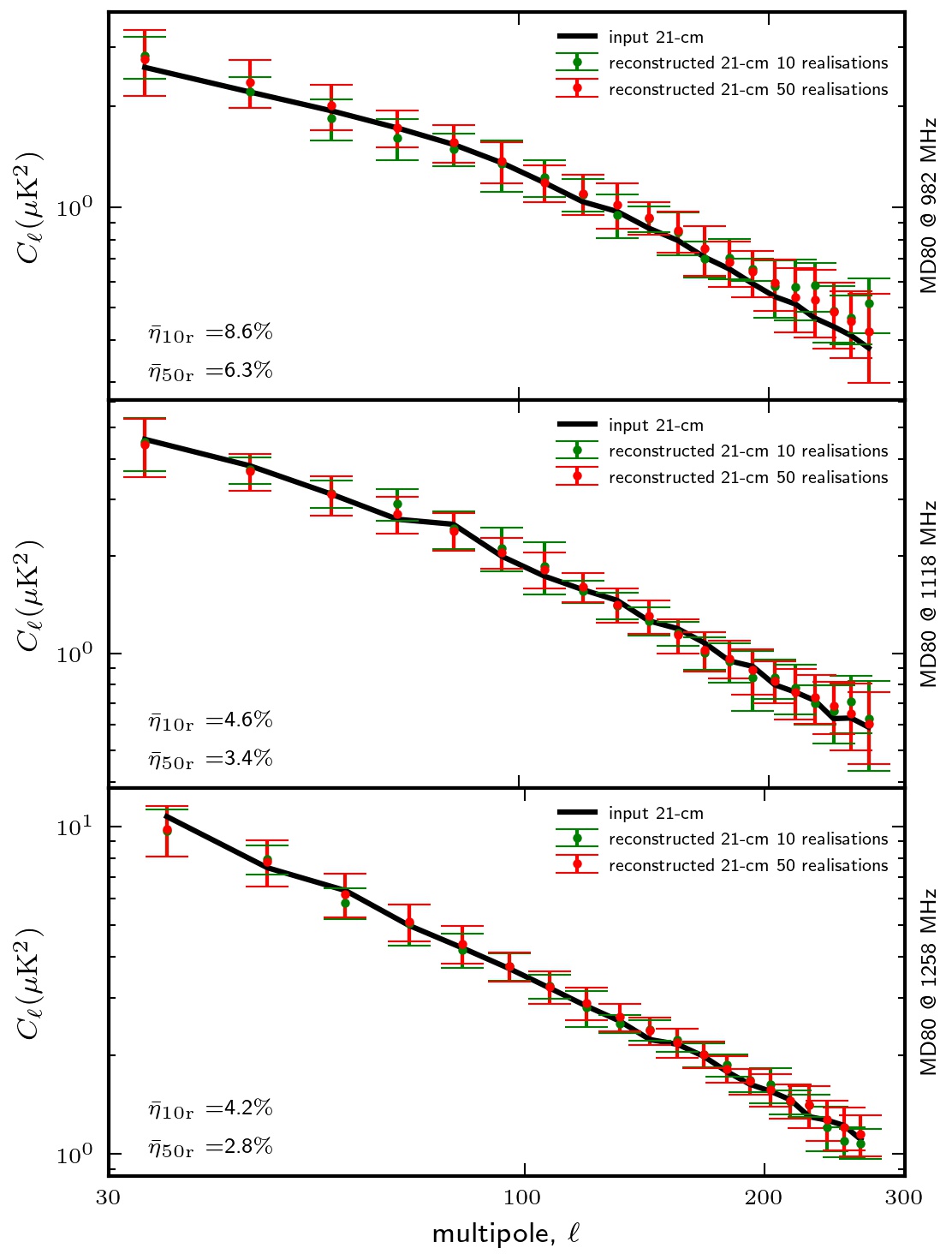}
\caption{Real (black) and reconstructed (green for $N_{\mathrm{realis}}=10$ and red for $N_{\mathrm{realis}}=50$) 21 cm signal all-sky power spectra, with respective error bars for the MD80 case (see Table \ref{tab:simulations_set}). The power spectra plotted refers to the channels centerd at 982 MHz ($Top$), 1118 MHz ($Center$) and 1258 MHz ($Bottom$).}
\label{fig:21cm_power_spectra_10r_and_50r}
\end{figure*}

To make our final estimate of the 21 cm power spectrum, we repeat for $N_{\mathrm{realis}}$ times the entire component separation procedure presented so far, considering different $N_{\mathrm{realis}}$ realizations of the {\tt GNILC} input maps (different realizations of \hi\ and noise plus foregrounds). Each debiasing procedure was performed with $N_{\mathrm{realis}}$ independent realizations of \hi\ and noise. Finally, we took the debiased power spectra and calculated the mean and the standard deviations (error bars). We did all this procedure for $N_{\mathrm{realis}}=10$ and $N_{\mathrm{realis}}=50$, using the data in the MD80 configuration. Figure \ref{fig:21cm_power_spectra_10r_and_50r} shows the 21 cm all-sky real and estimated power spectra for two different $N_{\mathrm{realis}}$ and in three different frequency channels. It can be observed that the spectrum relative to $N_{\mathrm{realis}}=50$ is closer to the real power spectrum in all the frequencies presented. The best result is obtained for $N_{\mathrm{realis}}=50$ at 1258 MHz, where the average power spectrum reconstruction error is $\bar{\eta}_{\mathrm{50r}}= 2.8\%$.

%\begin{figure*}
%\centering
%\includegraphics[scale=1.5]{figures/21cm_power_spectra_10r_and_50r.jpg}
%\caption{Real (black) and reconstructed (green for $N_{\mathrm{realis}}=10$ and red for $N_{\mathrm{realis}}=50$) 21 cm signal all-sky power spectra, with respective error bars for the MD80 case (see Table \ref{tab:simulations_set}). The power spectra plotted refers to the channels centerd at 982 MHz ($Top$), 1118 MHz ($Center$) and 1258 MHz ($Bottom$).}
%\label{fig:21cm_power_spectra_10r_and_50r}
%\end{figure*}

It is expected that a better estimate of the 21 cm power spectrum can be made with a larger $N_{\mathrm{realis}}$. However, due to our available computational capacity, we limit the number of realizations to $N_{\mathrm{realis}}=50$. The complete set of simulations for the MD80 configuration took two and a half months to be ready. For this we used 56 cores of the processor
Intel Xeon Gold 5120 2.20 GHz and 512 GB of RAM.

\section{Conclusions}
\label{sec:conclusions}

This paper presents two new results relevant to the BINGO data analysis and hardware configuration before commissioning time. The component separation procedure described here is based on {\tt GNILC}, as some previous papers from the collaboration  \citep{Liccardo:2020_mission-simulations, Fornazier:2020} and is  divided in two steps:
first, we recovered the \hi\ $+$ thermal noise maps applying {\tt GNILC} to the BINGO simulated data; second, we reconstructed the noiseless 21 cm power spectra passing the first step results through a debiasing procedure. For our analysis we used as default the BINGO project baseline configuration, MD30, and the number of \hi\ and noise realisations used in debiasing, $N_{\mathrm{realis}}=10$.

The first result points that {\tt GNILC} is robust against different foregrounds models. We tested two synchrotron models, finding that the reconstructed \hi\ plus noise maps in both cases have no significant differences in the pixel domain. In the harmonic domain, after the debiasing procedure, we obtained the mean \hi\ power spectra reconstruction error over all frequency channels of $7.2\%$, for the MD30 case, and of $6.4\%$ for the GD30 case, in the $30 \leq \ell \leq 270$ interval. The difference between foreground residuals of the two models is evident for $\ell <100$. This is due to the {\tt GNILC} ability to adapt the foregrounds removal not only to different regions of the sky, but also to different angular scales.

We also confirmed, as expected, that an extra channel with a simulated sky from CBASS experiment improves our 21 cm power spectrum reconstruction. Considering the MD30 configuration, the mean error in the power spectrum reconstruction decreases from 7.2\%, without CBASS information, to 5.7\% when we include the CBASS 5 GHz channel in the simulated data (MD30$+$CBASS case).

The second important result is a near optimal reconstruction of the \hi\ signal in a 80 frequency channel configuration for BINGO experiment. In this case, we obtained a mean error in our power spectrum reconstruction of 3\% in the multipole interval of $30 \leq \ell \leq 270$. This result was obtained by comparing the component separation results with simulated data with different numbers of frequency bins. These foregrounds removal runs indicated a stabilization trend in the reduction of the 21 cm signal estimation error with the increase in the number of channels. More precisely, we found that the recovery quality of the \hi\ signal starts stabilizing at $N_{\mathrm{ch}} = 60$ to $N_{\mathrm{ch}} = 80$, suggesting that it would not be worth using more than 100 redshift bins in the component separation process. The definition of the optimal number of frequency (redshift) bins will have a significant impact in the hardware configuration, helping to define the number of binning channels to preserve information from the raw data, $< 1 ~ms$ sampling mode.

We also tested the effect of the increase in the number of realizations of \hi\ and noise, used in the debiasing procedure, on the quality of the 21 cm signal estimate. The mean power spectrum reconstruction error, calculated over all channels, obtained with $N_{\mathrm{realis}}=10$ is  $\bar{\eta}=4.7\%$ and with $N_{\mathrm{realis}}=50$ is $\bar{\eta}=3.0\%$. These results indicate, as expected, that a greater number of \hi\ and noise realizations allows a better estimation and correction of additive and multiplicative biases of the {\tt GNILC} output power spectra. For an optimal choice of $N_{\mathrm{realis}}$ a more detailed analysis is necessary.

Finally, we repeated the entire component separation process for 10 and 50 different realizations of the BINGO simulated input data (sky $+$ noise), in the MD80 configuration, and obtained our final estimates of the reconstructed 21 cm spectra. We compared these results with the power spectrum of an independent 21 cm full-sky realization, considered here as the real map of \hi. Our estimated spectra are in good agreement with the input ones (independent realization). As expected, with 50 realizations it is possible to obtain a more accurate reconstruction of the \hi\ signal than with 10 realizations. The mean reconstruction error calculated in the range $30\leq\ell\leq270$ is equal to 6.3\%, 3.4\% and 2.8\% for channels centered at 982 MHz, 1118 MHz and 1258 MHz. 

Our results indicate that the debiasing procedure described in this work should work efficiently in BINGO data analysis pipeline. This suggests that, despite recovering the signal with good efficiency in the harmonic space covered by BINGO, a more detailed analysis of the debiasing procedure should be carried out in future works, particularly because we did not include the $1/f$ contribution to the overall noise in this analysis.

\bibliographystyle{aa}
\typeout{}
\bibliography{references} 

\begin{thebibliography}{50}
\expandafter\ifx\csname natexlab\endcsname\relax\def\natexlab#1{#1}\fi

\bibitem[{Abdalla {et~al.}(2021{\natexlab{a}})}]{Abdalla:2021_BINGO-project}
Abdalla, E. {et~al.} 2021{\natexlab{a}}, arXiv reprint:astro-ph
  [\eprint[arXiv]{2107.01633}]

\bibitem[{Abdalla
  {et~al.}(2021{\natexlab{b}})}]{2021/abdalla_BINGO-optical_design}
Abdalla, F.~B. {et~al.} 2021{\natexlab{b}}, arXiv reprint:astro-ph
  [\eprint[arXiv]{2107.01635}]

\bibitem[{Akaike(1974)}]{Akaike1974}
Akaike, H. 1974, IEEE transactions on automatic control, 19, 716

\bibitem[{Alonso {et~al.}(2019)Alonso, Sanchez, \& Slosar}]{Alonso:2019}
Alonso, D., Sanchez, J., \& Slosar, A. 2019, Mon. Not. Roy. Astron. Soc., 484,
  4127

\bibitem[{Ansari \& Magneville(2010)}]{ansari2010partial}
Ansari, R. \& Magneville, C. 2010, Monthly Notices of the Royal Astronomical
  Society, 405, 1421

\bibitem[{{Baccigalupi} {et~al.}(2004){Baccigalupi}, {Perrotta}, {de Zotti},
  {Smoot}, {Burigana}, {Maino}, {Bedini}, \& {Salerno}}]{Baccigalupi:2004}
{Baccigalupi}, C., {Perrotta}, F., {de Zotti}, G., {et~al.} 2004, \mnras, 354,
  55

\bibitem[{{Basak} \& {Delabrouille}(2012)}]{Basak:2012}
{Basak}, S. \& {Delabrouille}, J. 2012, Mon.\ Not.\ R.\ Astron.\ Soc., 419,
  1163

\bibitem[{{Basak} \& {Delabrouille}(2013)}]{Basak:2013}
{Basak}, S. \& {Delabrouille}, J. 2013, \mnras, 435, 18

\bibitem[{{Bennett} {et~al.}(2003){Bennett}, {Halpern}, {Hinshaw}, {Jarosik},
  {Kogut}, {Limon}, {Meyer}, {Page}, {Spergel}, {Tucker}, {Wollack}, {Wright},
  {Barnes}, {Greason}, {Hill}, {Komatsu}, {Nolta}, {Odegard}, {Peiris},
  {Verde}, \& {Weiland}}]{Bennett:2003}
{Bennett}, C.~L., {Halpern}, M., {Hinshaw}, G., {et~al.} 2003, \apjs, 148, 1

\bibitem[{{Betoule} {et~al.}(2009){Betoule}, {Pierpaoli}, {Delabrouille}, {Le
  Jeune}, \& {Cardoso}}]{Betoule:2009}
{Betoule}, M., {Pierpaoli}, E., {Delabrouille}, J., {Le Jeune}, M., \&
  {Cardoso}, J.~F. 2009, \aap, 503, 691

\bibitem[{{Bunn} {et~al.}(1994){Bunn}, {Fisher}, {Hoffman}, {Lahav}, {Silk}, \&
  {Zaroubi}}]{Bunn:1994}
{Bunn}, E.~F., {Fisher}, K.~B., {Hoffman}, Y., {et~al.} 1994, \apjl, 432, L75

\bibitem[{{Delabrouille} {et~al.}(2013){Delabrouille}, {Betoule}, {Melin},
  {Miville-Desch{\^e}nes}, {Gonzalez-Nuevo}, {Le Jeune}, {Castex}, {de Zotti},
  {Basak}, {Ashdown}, {Aumont}, {Baccigalupi}, {Banday}, {Bernard}, {Bouchet},
  {Clements}, {da Silva}, {Dickinson}, {Dodu}, {Dolag}, {Elsner}, {Fauvet},
  {Fa{\"y}}, {Giardino}, {Leach}, {Lesgourgues}, {Liguori},
  {Mac{\'\i}as-P{\'e}rez}, {Massardi}, {Matarrese}, {Mazzotta}, {Montier},
  {Mottet}, {Paladini}, {Partridge}, {Piffaretti}, {Prezeau}, {Prunet},
  {Ricciardi}, {Roman}, {Schaefer}, \& {Toffolatti}}]{Delabrouille:2013}
{Delabrouille}, J., {Betoule}, M., {Melin}, J.~B., {et~al.} 2013, \aap, 553,
  A96

\bibitem[{{Delabrouille} \& {Cardoso}(2009)}]{Delabrouille:2007}
{Delabrouille}, J. \& {Cardoso}, J.~F. 2009, in Data Analysis in Cosmology, ed.
  V.~J. {Mart{\'\i}nez}, E.~{Saar}, E.~{Mart{\'\i}nez-Gonz{\'a}lez}, \& M.~J.
  {Pons-Border{\'\i}a}, Vol. 665 (Springer), 159--205

\bibitem[{{Delabrouille} {et~al.}(2009){Delabrouille}, {Cardoso}, {Le Jeune},
  {Betoule}, {Fay}, \& {Guilloux}}]{Delabrouille:2009}
{Delabrouille}, J., {Cardoso}, J.~F., {Le Jeune}, M., {et~al.} 2009, \aap, 493,
  835

\bibitem[{{Delabrouille} {et~al.}(2003){Delabrouille}, {Cardoso}, \&
  {Patanchon}}]{Delabrouille:2003}
{Delabrouille}, J., {Cardoso}, J.~F., \& {Patanchon}, G. 2003, \mnras, 346,
  1089

\bibitem[{{Dragone}(1978)}]{Dragone:1978}
{Dragone}, C. 1978, AT T Technical Journal, 57, 2663

\bibitem[{{Eriksen} {et~al.}(2004){Eriksen}, {Banday}, {G{\'o}rski}, \&
  {Lilje}}]{Eriksen:2004}
{Eriksen}, H.~K., {Banday}, A.~J., {G{\'o}rski}, K.~M., \& {Lilje}, P.~B. 2004,
  \apj, 612, 633

\bibitem[{Eriksen {et~al.}(2008)Eriksen, Jewell, Dickinson, Banday,
  G{\'{o}}rski, \& Lawrence}]{Eriksen:2008}
Eriksen, H.~K., Jewell, J.~B., Dickinson, C., {et~al.} 2008, Astrophys.\ J.,
  676, 10

\bibitem[{Fornazier {et~al.}(2021)}]{Fornazier:2020}
Fornazier, K. {et~al.} 2021, submitted to A\%A

\bibitem[{{Giardino} {et~al.}(2002){Giardino}, {Banday}, {G{\'o}rski},
  {Bennett}, {Jonas}, \& {Tauber}}]{Giardino:2002}
{Giardino}, G., {Banday}, A.~J., {G{\'o}rski}, K.~M., {et~al.} 2002, \aap, 387,
  82

\bibitem[{{G{\'o}rski} {et~al.}(2005){G{\'o}rski}, {Hivon}, {Banday},
  {Wandelt}, {Hansen}, {Reinecke}, \& {Bartelmann}}]{Gorski:2005}
{G{\'o}rski}, K.~M., {Hivon}, E., {Banday}, A.~J., {et~al.} 2005, Astrophys.\
  J., 622, 759

\bibitem[{{Haslam} {et~al.}(1982){Haslam}, {Salter}, {Stoffel}, \&
  {Wilson}}]{Haslam:1982}
{Haslam}, C.~G.~T., {Salter}, C.~J., {Stoffel}, H., \& {Wilson}, W.~E. 1982,
  Astron.\ Astrophys. \ Suppl., 47, 1

\bibitem[{{Jewell} {et~al.}(2004){Jewell}, {Levin}, \&
  {Anderson}}]{Jewell:2004}
{Jewell}, J., {Levin}, S., \& {Anderson}, C.~H. 2004, \apj, 609, 1

\bibitem[{Jonas {et~al.}(1998)Jonas, Baart, \& Nicolson}]{jonas1998rhodes}
Jonas, J.~L., Baart, E.~E., \& Nicolson, G.~D. 1998, Monthly Notices of the
  Royal Astronomical Society, 297, 977

\bibitem[{Jones {et~al.}(2018)Jones, Taylor, Aich, Copley, Chiang, Davis,
  Dickinson, Grumitt, Hafez, Heilgendorff,
  {et~al.}}]{cbass:2018_design_and_capabilities}
Jones, M.~E., Taylor, A.~C., Aich, M., {et~al.} 2018, Monthly Notices of the
  Royal Astronomical Society, 480, 3224

\bibitem[{{Larson} {et~al.}(2007){Larson}, {Eriksen}, {Wandelt}, {G{\'o}rski},
  {Huey}, {Jewell}, \& {O'Dwyer}}]{Larson:2007}
{Larson}, D.~L., {Eriksen}, H.~K., {Wandelt}, B.~D., {et~al.} 2007, \apj, 656,
  653

\bibitem[{Liccardo {et~al.}(2021)}]{Liccardo:2020_mission-simulations}
Liccardo, V. {et~al.} 2021, arXiv reprint:astro-ph [\eprint[arXiv]{2107.01636}]

\bibitem[{Loureiro {et~al.}(2019)}]{Loureiro:2018qva}
Loureiro, A. {et~al.} 2019, Mon. Not. Roy. Astron. Soc., 485, 326

\bibitem[{{Maino} {et~al.}(2003){Maino}, {Banday}, {Baccigalupi}, {Perrotta},
  \& {G{\'o}rski}}]{Maino:2003}
{Maino}, D., {Banday}, A.~J., {Baccigalupi}, C., {Perrotta}, F., \&
  {G{\'o}rski}, K.~M. 2003, \mnras, 344, 544

\bibitem[{{Maino} {et~al.}(2002){Maino}, {Farusi}, {Baccigalupi}, {Perrotta},
  {Banday}, {Bedini}, {Burigana}, {De Zotti}, {G{\'o}rski}, \&
  {Salerno}}]{Maino:2002b}
{Maino}, D., {Farusi}, A., {Baccigalupi}, C., {et~al.} 2002, \mnras, 334, 53

\bibitem[{McLeod {et~al.}(2017)McLeod, Balan, \& Abdalla}]{mcleod2017joint}
McLeod, M., Balan, S.~T., \& Abdalla, F.~B. 2017, Monthly Notices of the Royal
  Astronomical Society, 466, 3558

\bibitem[{{Miville-Desch{\^e}nes} {et~al.}(2008){Miville-Desch{\^e}nes},
  {Ysard}, {Lavabre}, {Ponthieu}, {Mac{\'\i}as-P{\'e}rez}, {Aumont}, \&
  {Bernard}}]{Miville08}
{Miville-Desch{\^e}nes}, M.~A., {Ysard}, N., {Lavabre}, A., {et~al.} 2008,
  Astron.\ Astrophys., 490, 1093

\bibitem[{Murtagh \& Heck(1987)}]{Murtagh:1987_PCA}
Murtagh, F. \& Heck, A. 1987, Multivariate Data Analysis, Astrophys. Sp. Sc.
  Lib. 131

\bibitem[{{Olivari} {et~al.}(2016){Olivari}, {Remazeilles}, \&
  {Dickinson}}]{Olivari:2016}
{Olivari}, L.~C., {Remazeilles}, M., \& {Dickinson}, C. 2016, Mon.\ Not.\ R.\
  Astron.\ Soc., 456, 2749

\bibitem[{{Patanchon} {et~al.}(2005){Patanchon}, {Cardoso}, {Delabrouille}, \&
  {Vielva}}]{Patanchon:2005}
{Patanchon}, G., {Cardoso}, J.~F., {Delabrouille}, J., \& {Vielva}, P. 2005,
  \mnras, 364, 1185

\bibitem[{{Peel} {et~al.}(2019){Peel}, {Wuensche}, {Abdalla}, {Ant{\'o}n},
  {Barosi}, {Browne}, {Caldas}, {Dickinson}, {Fornazier}, {Monstein},
  {Strauss}, {Tancredi}, \& {Villela}}]{Peel:2019}
{Peel}, M.~W., {Wuensche}, C.~A., {Abdalla}, E., {et~al.} 2019, Journal of
  Astronomical Instrumentation, 8, 1940005

\bibitem[{{Peterson} {et~al.}(2006){Peterson}, {Bandura}, \&
  {Pen}}]{Peterson:2006}
{Peterson}, J.~B., {Bandura}, K., \& {Pen}, U.~L. 2006, arXiv reprint:astro-ph,
  0606104

\bibitem[{{Planck Collaboration} {et~al.}(2016){Planck Collaboration},
  {Aghanim}, {Ashdown}, {Aumont}, {Baccigalupi}, {Ballardini}, {Banday},
  {Barreiro}, {Bartolo}, {Basak}, {Benabed}, {Bernard}, {Bersanelli},
  {Bielewicz}, {Bonavera}, {Bond}, {Borrill}, {Bouchet}, {Boulanger},
  {Burigana}, {Calabrese}, {Cardoso}, {Carron}, {Chiang}, {Colombo}, {Comis},
  {Couchot}, {Coulais}, {Crill}, {Curto}, {Cuttaia}, {de Bernardis}, {de
  Zotti}, {Delabrouille}, {Di Valentino}, {Dickinson}, {Diego}, {Dor{\'e}},
  {Douspis}, {Ducout}, {Dupac}, {Dusini}, {Elsner}, {En{\ss}lin}, {Eriksen},
  {Falgarone}, {Fantaye}, {Finelli}, {Forastieri}, {Frailis}, {Fraisse},
  {Franceschi}, {Frolov}, {Galeotta}, {Galli}, {Ganga}, {G{\'e}nova-Santos},
  {Gerbino}, {Ghosh}, {Giraud-H{\'e}raud}, {Gonz{\'a}lez-Nuevo}, {G{\'o}rski},
  {Gruppuso}, {Gudmundsson}, {Hansen}, {Helou}, {Henrot-Versill{\'e}},
  {Herranz}, {Hivon}, {Huang}, {Jaffe}, {Jones}, {Keih{\"a}nen}, {Keskitalo},
  {Kiiveri}, {Kisner}, {Krachmalnicoff}, {Kunz}, {Kurki-Suonio}, {Lamarre},
  {Langer}, {Lasenby}, {Lattanzi}, {Lawrence}, {Le Jeune}, {Levrier}, {Lilje},
  {Lilley}, {Lindholm}, {L{\'o}pez-Caniego}, {Ma}, {Mac{\'\i}as-P{\'e}rez},
  {Maggio}, {Maino}, {Mandolesi}, {Mangilli}, {Maris}, {Martin},
  {Mart{\'\i}nez-Gonz{\'a}lez}, {Matarrese}, {Mauri}, {McEwen}, {Melchiorri},
  {Mennella}, {Migliaccio}, {Miville-Desch{\^e}nes}, {Molinari}, {Moneti},
  {Montier}, {Morgante}, {Moss}, {Natoli}, {Oxborrow}, {Pagano}, {Paoletti},
  {Patanchon}, {Perdereau}, {Perotto}, {Pettorino}, {Piacentini},
  {Plaszczynski}, {Polastri}, {Polenta}, {Puget}, {Rachen}, {Racine},
  {Reinecke}, {Remazeilles}, {Renzi}, {Rocha}, {Rosset}, {Rossetti}, {Roudier},
  {Rubi{\~n}o-Mart{\'\i}n}, {Ruiz-Granados}, {Salvati}, {Sandri}, {Savelainen},
  {Scott}, {Sirignano}, {Sirri}, {Soler}, {Spencer}, {Suur-Uski}, {Tauber},
  {Tavagnacco}, {Tenti}, {Toffolatti}, {Tomasi}, {Tristram}, {Trombetti},
  {Valiviita}, {Van Tent}, {Vielva}, {Villa}, {Vittorio}, {Wandelt}, {Wehus},
  {Zacchei}, \& {Zonca}}]{Planck2016:dustemission}
{Planck Collaboration}, {Aghanim}, N., {Ashdown}, M., {et~al.} 2016, \aap, 596,
  A109

\bibitem[{{Reich} \& {Reich}(1986)}]{Reich:1986}
{Reich}, P. \& {Reich}, W. 1986, Astron.\ Astrophys. \ Suppl., 63, 205

\bibitem[{{Remazeilles} {et~al.}(2011{\natexlab{a}}){Remazeilles},
  {Delabrouille}, \& {Cardoso}}]{Remazeilles:2011b}
{Remazeilles}, M., {Delabrouille}, J., \& {Cardoso}, J.-F. 2011{\natexlab{a}},
  Mon.\ Not.\ R.\ Astron.\ Soc., 410, 2481

\bibitem[{{Remazeilles} {et~al.}(2011{\natexlab{b}}){Remazeilles},
  {Delabrouille}, \& {Cardoso}}]{Remazeilles:2011}
{Remazeilles}, M., {Delabrouille}, J., \& {Cardoso}, J.-F. 2011{\natexlab{b}},
  \mnras, 418, 467

\bibitem[{{Remazeilles} {et~al.}(2015){Remazeilles}, {Dickinson}, {Banday},
  {Bigot-Sazy}, \& {Ghosh}}]{Remazeilles:2015}
{Remazeilles}, M., {Dickinson}, C., {Banday}, A.~J., {Bigot-Sazy}, M.~A., \&
  {Ghosh}, T. 2015, Mon.\ Not.\ R.\ Astron.\ Soc., 451, 4311

\bibitem[{{Saha} {et~al.}(2006){Saha}, {Jain}, \& {Souradeep}}]{Saha:2006}
{Saha}, R., {Jain}, P., \& {Souradeep}, T. 2006, \apjl, 645, L89

\bibitem[{{Tegmark} {et~al.}(2003){Tegmark}, {de Oliveira-Costa}, \&
  {Hamilton}}]{Tegmark:2003}
{Tegmark}, M., {de Oliveira-Costa}, A., \& {Hamilton}, A.~J. 2003, \prd, 68,
  123523

\bibitem[{{Tegmark} \& {Efstathiou}(1996)}]{Tegmark:1996}
{Tegmark}, M. \& {Efstathiou}, G. 1996, \mnras, 281, 1297

\bibitem[{{Wandelt} {et~al.}(2004){Wandelt}, {Larson}, \&
  {Lakshminarayanan}}]{Wandelt:2004}
{Wandelt}, B.~D., {Larson}, D.~L., \& {Lakshminarayanan}, A. 2004, \prd, 70,
  083511

\bibitem[{{Wolz} {et~al.}(2015){Wolz}, {Blake}, {Abdalla}, {Anderson}, {Chang},
  {Li}, {Masui}, {Switzer}, {Pen}, {Voytek}, \& {Yadav}}]{Wolz:2015}
{Wolz}, L., {Blake}, C., {Abdalla}, F.~B., {et~al.} 2015, arXiv e-prints,
  arXiv:1510.05453

\bibitem[{{Wuensche} {et~al.}(2020){Wuensche}, {Reitano}, {Peel}, {Browne},
  {Maffei}, {Abdalla}, {Radcliffe}, {Abdalla}, {Barosi}, {Liccardo}, {Mericia},
  {Pisano}, {Strauss}, {Vieira}, {Villela}, \& {Wang}}]{Wuensche:2020}
{Wuensche}, C.~A., {Reitano}, L., {Peel}, M.~W., {et~al.} 2020, Experimental
  Astronomy, 50, 125

\bibitem[{Wuensche {et~al.}(2021)}]{Wuensche:2021_instrument-description}
Wuensche, C.~A. {et~al.} 2021, arXiv reprint:astro-ph
  [\eprint[arXiv]{2107.01634}]

\bibitem[{{Xavier} {et~al.}(2016){Xavier}, {Abdalla}, \& {Joachimi}}]{Xavier16}
{Xavier}, H.~S., {Abdalla}, F.~B., \& {Joachimi}, B. 2016, Mon.\ Not.\ R.\
  Astron.\ Soc., 459, 3693

\end{thebibliography}

\begin{acknowledgements}
The BINGO project is supported by FAPESP grant 2014/07885-0. E.J.M. acknowledges CAPES for the Ph.D. fellowship. C.A.W. acknowledges a CNPq grant 313597/2014-6.  
T.V. acknowledges  CNPq  Grant  308876/2014-8. 
%E.A. gracefully acknowledges the CNPq support. 
L.S. is supported by the National Key R\&D Program of China (2020YFC2201600) and NSFC grant 12150610459.
The authors thank the HEALPix creators for the HEALPix package (http://healpix.sourceforge.net) \citep{Gorski:2005}.
\end{acknowledgements}

\end{document}